\documentclass[10pt]{article}
\usepackage{epsfig}
\def\mapbelow#1{\smash{\mathop{\longrightarrow}\limits_{#1}}}

\newcommand{\lab}{\label}

\newcommand{\bc}{\begin{center}}
\newcommand{\ec}{\end{center}}
\newcommand{\be}{\begin{equation}}
\newcommand{\ee}{\end{equation}}
\newcommand{\bea}{\begin{eqnarray}}
\newcommand{\eea}{\end{eqnarray}}
\newcommand{\bs}{\begin{subequations}}
\newcommand{\es}{\end{subequations}}
\newcommand{\beq}{\begin{eqalignno}}
\newcommand{\eeq}{\end{eqalignno}}
\def\lab{\label}
\begin{document}

\normalsize

\author{Walter J. Freeman$^\dagger $ and Giuseppe Vitiello$^{\dagger\dagger
}$\\ \\
$^{\dagger}$Department of Molecular and Cell Biology,\\ University
of California, Berkeley CA 94720-3206 USA\\ dfreeman@berkeley.edu
- http://sulcus.berkeley.edu\\ \\
$^{\dagger \dagger }$Dipartimento di Fisica ``E.R. Caianiello'',
INFN and INFM,\\ Universit\'a degli Studi di Salerno, 84100
Salerno,
Italia\\
vitiello@sa.infn.it -
http://www.sa.infn.it/giuseppe.vitiello/vitiello/}

%Vitiello$^{\dagger\dagger }$\footnote{communicating author:
%G.~Vitiello,
%vitiello@sa.infn.it}~, A. Widom\\ \\

\title{Nonlinear brain dynamics as macroscopic manifestation\\ of underlying
many-body field dynamics}
\date{}

%\begin{document}
\maketitle

\vskip .7cm
\centerline{Abstract}
\medskip
\par \noindent
{Neural activity patterns related to behavior occur at many scales
in time and space from the atomic and molecular to the whole
brain. Patterns form through interactions in both directions, so
that the impact of transmitter molecule release can be analyzed
upwardly through synapses, dendrites, neurons, populations and
brain systems to behavior, and control of that release can be
described step-wise through top-down transformations. Here we
explore the feasibility of interpreting neurophysiological data in
the context of many-body physics by using tools that physicists
have devised to analyze comparable hierarchies in other fields of
science. We focus on a mesoscopic level that offers a multi-step
pathway between the microscopic functions of neurons and the
macroscopic functions of brain systems revealed by hemodynamic
imaging. We use electroencephalographic (EEG) records collected
from high-density electrode arrays fixed on the epidural surfaces
of primary sensory and limbic areas in rabbits and cats trained to
discriminate conditioned stimuli (CS) in the various modalities.
High temporal resolution of EEG signals with the Hilbert transform
gives evidence for diverse intermittent spatial patterns of
amplitude (AM) and phase modulations (PM) of carrier waves that
repeatedly re-synchronize in the beta and gamma ranges at near
zero time lags over long distances. The dominant mechanism for
neural interactions by axodendritic synaptic transmission should
impose distance-dependent delays on the EEG oscillations owing to
finite propagation velocities. It does not. EEGs instead show
evidence for anomalous dispersion: the existence in neural
populations of a low velocity range of information and energy
transfers, and a high velocity range of the spread of phase
transitions. This distinction labels the phenomenon but does not
explain it. In this report we explore the analysis of these
phenomena using concepts of energy dissipation, the maintenance by
cortex of multiple ground states corresponding to AM patterns, and
the exclusive selection by spontaneous breakdown of symmetry (SBS)
of single states in sequences.}

\bigskip
\medskip
\par \noindent


%PACS: 02.40.Gh, 11.10.Ef
%\vfill
%\eject
\newpage

\section{Introduction}

~~~~~{\bf 1a. Overview: Phenomena observed in brain dynamics and
explained using tools of physics}

\vspace{0.3cm}

Classical physics has provided a strong foundation for
understanding brain function through measuring brain activity,
modeling the functional connectivity of networks of neurons with
algebraic matrices, and modeling the dynamics of neurons and
neural populations with sets of coupled differential equations
(Freeman, 1975/2004, 2000, Appendix A). The areas of physics that
have been most fruitful have been Newtonian mechanics, the theory
of the potential, optics, thermodynamics, and statistical
mechanics. These tools enabled recognition and documentation of
the physical states of brains; the structures and dynamics of
neurons; the operations of membranes and organelles that generate
and channel electric currents; and the molecular and ionic
carriers that implement the neural machineries of electrogenesis
and learning. They support description of brain functions at
several levels of complexity through measuring neural activity in
the brains of animal and human subjects engaged in behavioral
exchanges with their environments. In these brain states a salient
property has been the coordinated oscillations of populations of
neurons that were changing rapidly in concert with changes in the
environment - more specifically, with the evolution of the
meaningful relationships between each subject and its environment
that were established and maintained by the action-perception
cycle (Freeman, 2001; Vitiello, 2001).

Most experimental neurobiologists and neural theorists have
focused on sensorimotor functions and their adaptations through
various forms of learning and memory (e.g., Singer and Gray, 1998;
Orlovskii, Deliagina and Grillner, 1999; Hawkins, 2004). Principal
reliance has been placed on measurements of the rates and
intervals of trains of action potentials of small numbers of
neurons and modeling neural interactions with discrete networks of
simulated neurons. In our work we have focused on the fields of
potential established by dendritic currents of populations of
neurons, which are often called local field potentials (LFP) when
recorded in the brain, electrocorticograms (ECoG) when taken from
the cortical surface, and electroencephalograms (EEG) when at the
scalp. Here we use "EEG" irrespective of recording site. Models
for these oscillations have included linear filters of noise,
either white (Bullock, 1969; Elul, 1972; Robinson et al., 2004) or
1/f (Linkenkaer-Hansen et al., 2001; Hwa and Ferree, 2002;
Wakeling, 2004). Mechanisms have been modeled by negative feedback
among local populations (Basar, 1998), thalamocortical loops
(Andersen and Andersson, 1968; Steriade, 1997; Hoppensteadt and
Izhkevich, 1998; Miller and Schreiner, 2000), subcortical loops
(Houk, 2001), electrochemical reactions based in membrane
permeabilities (Traub et al., 1996; Whittington et al., 2000), and
resonant modes of wave mechanics (Nunez, 1981). Here we depend on
models with multiple types of nonlinear distributed feedback
(Freeman, 1975/2004).

The types of behavior we address included high-level
generalization, abstraction, and contextualization effected late
in the action-perception cycle of Piaget (1930), Merleau-Ponty
(1945/1962), and Maturana and Varela (1980). Our experimental
designs and modeling are based on the neuropsychological field
theories of Lashley (1929), K\"ohler (1940). and Pribram (1971).
Karl Lashley wrote: "Generalization [stimulus equivalence] is one
of the primitive basic functions of organized nervous tissue.  …
Here is the dilemma.  Nerve impulses are transmitted from cell to
cell through definite intercellular connections.  Yet all behavior
seems to be determined by masses of excitation. ... What sort of
nervous organization might be capable of responding to a pattern
of excitation without limited specialized paths of conduction? The
problem is almost universal in the activities of the nervous
system"  (1942, p. 306).  Wolfgang K\"ohler wrote: "Our present
knowledge of human perception leaves no doubt as to the general
form of any theory which is to do justice to such knowledge: a
theory of perception must be a field theory.  By this we mean that
the neural functions and processes with which the perceptual facts
are associated in each case are located in a continuous medium"
(1940, p. 55). Pribram (1971) proposed a holographic model to
explain psychological field data. Bartlett (1932) wrote: "...some
widely held views [of memory] have to be completely discarded, and
none more completely than that which treats recall as the
re-excitement in some way of fixed and changeless 'traces' " (p.
vi). We therefore sought to devise a neurodynamics of dendritic
potentials by which to synthesize contemporary unit data with
classical mid-20th century perceptual field theory into a new
framework.

Our analysis of electroencephalographic activity (EEG) has shown
that cortical activity does not change continuously with time but
by multiple spatial patterns in sequences during each perceptual
action that resemble cinematographic frames on multiple screens
(Freeman, Burke and Holmes, 2003; Freeman, 2004a, b). We identify
the source of these patterns with regions in the embedding medium
of neurons, the neocortical neuropil: the dense felt-work of
axons, dendrites, cell bodies, glia and capillaries forming a
superficial continuum $1- 3 ~mm$ in thickness over the entire
extent of each cerebral hemisphere in mammals. The carrier waves
of the patterned activity in frames come in at least two ranges
that we identify with beta ($12-30 ~Hz$) and gamma ($30-80 ~Hz$)
oscillations in animal EEG (Freeman, 2005a,b, 2006). The abrupt
change in dynamical state with each new frame is not readily
amenable to description either with classic integrodifferential
equations (e.g., Basar, 1998; Wright, Bourke and Chapman, 2000) or
the algebras of neural networks. We propose that each frame is
formed by a {\it phase transition} (Freeman, 2004b).

The initiation and maintenance of shared oscillations by this
phase transition requires rapid communication among neurons.
Several alternative mechanisms have been proposed as the agency
for widespread synchrony. These include dendritic loop current as
the chief agent for {\it intracellular} communication and the
axonal action potential as the chief agent for {\it intercellular}
communication. The propagating action potential is the mechanism
by which multicellular organisms greater in size than about a
millimeter overcome the limitations of diffusion by providing for
rapid communication over long distances without attenuation.
Diffusion of chemical transmitters substances remains the agent in
synaptic clefts and extracellular channels with lengths on the
order of hundreds of microns. Indeed the cable equation used to
describe passive loop currents also describes a 1-dimensional
diffusion process. Passive conduction suffices for distances less
than $1 ~mm$; neurons exceeding that range are well known to have
dendritic spike mechanisms that complement diffusion. Fuxe and
Agnati (1991) and Bach-y-Rita (1995) have compiled extensive
evidence for what they call {\it nonsynaptic diffusion
neurotransmission} to explain {\it volume transmission}. We
believe that the temporal precision and fineness of spatial
texture of synchronized cortical activity that we have documented
over distances of $mm$ to $cm$ (Freeman, 2005b) are incompatible
with the mechanism of long-range diffusion.

Communication by propagating action potentials imposes
distance-dependent delays in the onset of resynchronization during
a phase transition over an area of cortex. The delays are
measurable as distance-dependent phase lags at the various
frequencies of oscillation (Freeman, 2004b). However, the length
of most axons in cortex is a small fraction of observed distances
of long-range correlation, with the requirement for synaptic
renewal at each successive relay. Even the presence of relatively
sparse long axons, which provide for high velocity jumps to seed
areas over long distances creating {\it small-world} effects
(Watts and Strogatz, 1998; Kozma et al., 2005), cannot easily
explain the observed long-range correlations, which are maintained
despite continuous variations in transmission frequencies that are
apparent in aperiodic {\it chaotic} oscillations. Several groups
have proposed that zero-lag correlation between oscillations at
distant sites might be explained by reciprocal transmission at
equal velocities (K\"onig and Schillen, 1991; Roelfsma et al.,
1997; Wright, Alexander and Bourke, 2005).  We believe that the
evidence points instead to the classical physical description of
{\it anomalous dispersion} (Freeman, 1990), by which the rate of
propagation of phase transitions through cortex might exceed the
rates of propagation of energy and information. However, putting a
name on the phenomenon does not explain how zero-lag correlation
(Freeman, Ga\'al and Jornten, 2003) can emerge over distances of
many cm in less than $1/4$ cycle of oscillations in the beta and
gamma ranges ($12-80+ ~Hz$) in neocortex.

Some researchers have sought to explain zero-lag correlations with
processes other than axodendritic synaptic transmission. Both
electric fields and magnetic fields accompany neural loop
currents. Bullock (1959) showed in vitro that weak electric fields
modulate neural firing. However, the intensities of the
extracellular electric forces are $3$ orders of magnitude smaller
than the transmembrane potential differences. The electric
potential gradients of the EEG have been shown (Freeman and Baird,
1989) to be inadequate in vivo to account for the long-range of
the observed coherent activity, largely owing to the shunting
action of glia that reduce the fraction of extracellular dendritic
current penetrating adjacent neurons and minimize ephaptic
cross-talk among cortical neurons. Like the decay in diffusion
potential the fall of electric potential with distance is too
rapid, and the intensity of the Coulomb forces is too weak to
explain either the abruptness of phase transitions manifested in
EEGs or the entry of beta and gamma oscillations into synchrony
within the observed time windows of a few ms over distances of $1$
to $19 ~cm$ (Freeman and Barrie, 2000; Freeman, Ga\'al and
Jornten, 2003; Freeman and Burke, 2003; Freeman and Rogers, 2003;
Freeman, Burke and Holmes, 2003). Regarding the agency of magnetic
fields, the search in cortex for magnetic field receptors like
those in bees and birds serving navigation remains inconclusive
(Dunn et al., 1995). Moreover, there are no significant
electromagnetic fields (radio waves) in brains; EEG oscillations
are too slow and excessive in wavelength, and brain electric and
magnetic permeabilities differ $80:1$.

As a reasonable alternative we turn to the mathematical machinery
of many-body field theory that enables us to describe phase
transitions in distributed nonlinear media having innumerable
co-existing and overlapping ground states, actual and potential.
One might wonder about the necessity and the correctness of using
many-body field theory in treating brain dynamics. The common
belief is that, if physics has to be involved in the description
of brain dynamics, classical tools such as non-linear dynamics and
statistical mechanics should suffice.  However, many-body field
theory appears to us as the only existing theoretical tool capable
to explain the {\it dynamic origin} of long-range correlations,
their rapid and efficient formation and dissolution, their interim
stability in ground states, the multiplicity of coexisting and
possibly non-interfering ground states, their degree of ordering,
and their rich textures. It is historical fact that many-body
quantum field theory has been devised and constructed in past
decades exactly to understand features like ordered pattern
formation and phase transitions in condensed matter physics,
similar to those in the brains, that could not be understood in
classical physics.

The notion of coherent collective modes, which are macroscopic
features of quantum origin, has been used in many practical
applications (in solid state physics and laser physics, for
example). In a familiar crystal, a magnet, the ordered patterns
observed at room temperature, without recourse to low-temperature
superconductivity or superfluidity, are well known examples of
{\it macroscopic quantum systems} (Umezawa, 1993). The domain of
validity of the 'quantum' is not restricted to the microscopic
world. There are macroscopic features of classically behaving
systems, like the ones just mentioned, which cannot be explained
without recourse to the quantum dynamics. This does not mean that
the biochemistry, neurophysiological analysis and/or any other
classical tool of investigation might be superceded. Rather, it
means that brain studies made by continuing use of these
traditional classical tools might be enhanced by descriptions of
the underlying dynamics that will enable us to understand the
phenomenology.

One final remark is that quantum field theory (QFT) differs
drastically from quantum mechanics (QM) and must not be confused
with it. To explain such a difference goes outside the task of the
present paper. We only say that multiple, physically distinct,
ground states do not exist in QM, contrarily to what happens in
many-body field theory, and thus there cannot co-exist different
physical phases for a system in QM. The interested reader can find
more on this in Umezawa (1993) and  in Vitiello (2001) (see also
Umezawa and Vitiello, 1985).

\vspace{0.3cm}

{\bf 1b. The many-body model and the macroscopic collective field
of action}

\vspace{0.3cm}

Our field theoretic model leads to our view of the phase
transition as a condensation that is comparable to the formation
of fog and raindrops from water vapor, and that might serve to
model both the gamma and beta phase transitions. According to such
a model, originally proposed by Ricciardi and Umezawa (1967),
developed by others (Stuart et al., 1978; 1979; Jibu and Yasue;
1992; 1995), and extended to dissipative dynamics (Vitiello, 1995;
2001), the production of activity with long-range correlation in
the brain takes place through the mechanism of spontaneous
breakdown of symmetry (SBS), which has for decades been shown to
describe long-range correlation in condensed matter physics (see also  
Kelso, 1995; Haken, 1996; 2004; Bressler and Kelso,
2001 for a field theoretical approach to brain modeling). The
adoption of such a field theoretic approach enables us to model
the whole cerebral hemisphere and its hierarchy of components down
to the atomic level as a fully integrated macroscopic quantum
system, namely as a macroscopic system which is a quantum system
not in the trivial sense that it is made, like all existing
matter, by quantum components such as atoms and molecules, but, as
already mentioned, in the sense that some of its macroscopic
properties can best be described with recourse to quantum
dynamics.

A central concern in our attempt to apply many-body physics is
then expressed in the question:  What might be the {\it bridge}
between microscopic, atomic and molecular, units and the
macroscopic neural activity as we observe it? Traditionally, the
unit of neural activity is taken to be the action potential, the
dendritic postsynaptic potential (PSP), the chemical packet in the
synaptic vesicle, an extracellular diffusion field, or a gap
junction.  On the one hand, the neuron, cell body, synapse,
microtubule, vesicle, or other microscopic structures {\it are not
to be considered as quantum objects} in our analysis. The Planck
constant, $h$, is undeniably the unit of action at the atomic
scale and below, but it is not the decisive factor at the level of
neuronal populations. On the other hand, what appears to emerge
from our experiments is a {\it wave packet} (Freeman, 1975/2004,
2000) acting as a bridge from quantum dynamics at the atomic level
through the microscopic pulse trains of neurons to the macroscopic
properties of large populations of neurons. The wave packet we
refer to is a macroscopic collective field of action that has
measurable field properties: the phase, the amplitude, and their
spatial and temporal rates of change (gradients and frequencies)
at each point in time and space in the sustaining neuropil. Our
measurements of the durations, recurrence intervals and diameters
of neocortical EEG phase patterns have power-law distributions
with no detectable minima. The means and standard deviations (SD)
vary depending on the scale of measurement within the limits of
observation. The power spectral densities in both time ($PSD_{T}$)
and space ($PSD_{X}$) of EEGs from surface arrays likewise conform
to power-law distributions, $1/f^{a}$, with $a$ commonly near $2$.
These findings (Freeman, 2004a,b) and those of others (Braitenberg
and Sch\"uz, 1991; Linkenkaer-Hansen et al., 2001; Hwa and Ferree,
2002) suggest that the activity patterns generated by neocortical
neuropil might be scale-free (Wang and Chen, 2003; Freeman, 2005c)
with self-similarity in EEG patterns over distances ranging from
hypercolumns to an entire cerebral hemisphere, thus explaining the
similarity of neocortical dynamics in mammals differing in brain
size by 4 orders of magnitude, from mouse (Franken, Malafosse and
Tafti, 1998) to whale (Lyamin et al., 2002), which contrasts
strikingly with the relatively small range of size of avian,
reptilian and dinosaur brains lacking neocortex.

It is important again to stress that our wave packet is not to be
confused with the notion of wave packet describing probability
amplitudes in QM (the common denomination is only accidental). In
our field-theoretic approach, the wave packet is {\it a collective
mode that sustains a field of neural activity} (Freeman,
1975/2004), which gives rise to the observable fields of EEGs and
action potentials we record from arrays of electrodes. We propose
to explore the utility of modeling the wave packet observed in our
experiments with the Nambu-Goldstone boson mode in
Bose-Einstein condensates in the dissipative many-body model
(Vitiello, 1995, 2001).

Our paper is organized as follows. In Sections 2 and 3 we present
analysis of EEG observations. In Sections 4 and 5 we introduce the
concepts of SBS, order, and long-range correlation in the
many-body model. In Sections 6 we analyze the role of dissipation
in the observed phase transitions and related experimental
observations. We devote Sections 7-9 to exploring the implications
of our findings, and Section 10 to concluding remarks and
comments. Appendix A outlines K-fields and K-sets; Appendix B
summarizes the mathematical formalism of our many-body model.

Our talisman in this scientific endeavor is a human subject, who
may be imagined as a bright young woman with intractable epilepsy
(Freeman, Holmes et al., 2005). Her partial complex seizures
robbed her of consciousness and replaced intentional actions with
meaningless movements. Failure of medication led to neurosurgical
evaluation requiring placement of intracranial electrode arrays.
Her EEG data were transmitted without personal identification from
clinic to laboratory; her history and therapeutic outcome are
curtained by anonymity.

\section{Observation and measurement of multiple EEG signals at high
space-time resolution}

Two main features of our EEG data justify our proposal to use QFT
for explication: the textured patterns of amplitude modulation
(AM) in distinct frequency bands that are correlated with
categories of conditioned stimuli (CS), and the tight sequencing
of AM patterns in epochs that resemble cinematographic frames. We
propose to identify each spatial AM pattern with a ground state,
and its manner of onset as a SBS. The high spatial resolution
required to measure AM pattern textures came by designing 2-D
square high-density electrode arrays (typically $ 8 \times 8$) and
fixing them on the scalp or the epidural surface of cortical
areas. We determined the optimal interelectrode intervals by using
1-D curvilinear arrays to over-sample the spatial EEG
distributions with ultra-close spacing and the 1-D fast Fourier
transform (FFT) to calculate the spatial spectrum and fix a
spatial Nyquist frequency that would avoid both aliasing and
over-sampling. The high temporal resolution required for
sequencing came by applying the Hilbert transform to EEG signals
after band pass filtering. We optimized filter settings by
constructing tuning curves with the criterion of maximal
classification of spatial AM patterns with respect to CS. Unlike
the Fourier transform that decomposed an extended time series into
fixed frequency components, the Hilbert transform decomposed an
EEG signal into the analytic amplitude and the analytic phase at
each digitizing time step on each channel: a point in time and
space. Owing to the relative invariance of the extracellular
impedance (Freeman, 1975/2004) the square of the analytic
amplitude, $A^{2}(t)$, gave an estimate of the instantaneous power
expended by a local neighborhood of cortex in generating the ionic
currents underlying the EEG. The set of $n$ amplitudes from an
array of $n$ electrodes (typically $64$) defined a feature vector,
${\bf A}^{2}(t)$, at each time step for a set of $n$ points on the
cortical surface. The vector specified a point dynamic in {\it
brain state-space}, which we conceived as the collection of all
possible brain states, essentially infinite. Our measurement of n
EEG signals defined a finite n-dimensional subspace, so the point
specified by ${\bf A}^{2}(t)$ was unique to a spatial AM pattern
of the aperiodic carrier wave. Similar AM patterns formed a
cluster in n-space, and multiple patterns formed either multiple
clusters or trajectories with large Euclidean distances between
digitizing steps through n-space. When we measured a cluster by
its center of gravity and its $SD$ radius and verified its
behavioral correlate, then by any of several statistical
classification techniques we assigned a category of CS to that
cluster (Ohl, Scheich and Freeman, 2001; 2005; Kozma and Freeman,
2001; Freeman, 2005). A cluster with a verified behavioral
correlate denoted an {\it ordered AM pattern and an ordered wave
packet}.

The vector served also as our order parameter, because when the
trajectory of a sequence of points entered into a cluster, that
location in state space signified increased order from the
perspective of an intentional state of the brain, owing to the
correlation with a CS. The Euclidean distance between successive
points in n-space, $D_{e}(t)$, specified the rate of change in the
order parameter. We calculated $D_{e} = [{\bf A}^{2}(t) - {\bf
A}^{2}(t-1)]$ after frame normalization, as distinct from $\Delta
\underline{A}^{2} (t) = [\underline{A}^{2} (t) - \underline{A}^{2}
(t-1)]$ without normalization, the latter being the rate of change
in mean power. Typically $D_{e}(t)$ took large steps between
clusters, decreased to a low value when the trajectory entered a
cluster, and remained low for tens of $ms$ within a frame.
Therefore $D_{e}(t)$ served as a measure of the spatial AM pattern
stability as one property of a wave packet.

The analytic phase at each digitizing step and electrode specified
the {\it instantaneous frequency}, $\omega (t)$, in the designated
band as an increment in phase divided by an increment in time in
$rad/s$. Our measurements showed that typically the rate of change
in $\omega (t)$ was low in frames that coincided with low
$D_{e}(t)$ indicating stabilization of frequency as well as AM
pattern. Between frames $\omega (t)$ increased often several fold
or decreased even below zero in interframe breaks that repeated at
rates in the theta or alpha range of the EEG (Freeman, Burke and
Holmes, 2003). Such apparent discontinuities in phase were
designated {\it phase slip} (Pikovsky, Rosenblum and Kurths,
2001). In array recordings the phase slips tended to occur almost
simultaneously over the entire array. Therefore a useful index of
the temporal stability was the spatial standard deviation of
analytic phase differences, $SD_{X}(t)$. Peaks bracketed the
stabilized epochs and defined the beginning and end of wave
packets.

The commonality of the aperiodic carrier wave demonstrated by
visual inspection that the neural activity was synchronized across
a designated spectral range. Commonality was confirmed by showing
that the first component in principal components analysis
typically contained more than $95\%$ of the total variance. The
continual variation in the frequency impeded the use of phase to
measure synchronization; the EEG usually had featureless,
monotonous fluctuations with no periodic activity. The $1/f^{a}$
$PSD_{T}$ in log-log coordinates with $a \sim  2$ conformed to
Brownian motion (Freeman, Burke, Holmes and Vanhatalo, 2003) with
no stable peaks by which to define frequency or phase. A
quantitative index of synchrony independent of phase was devised
for arrays of aperiodic oscillatory EEG signals in a moving window
twice the wavelength of the center frequency of the pass band
required for the Hilbert transform: the ratio, $R_{e}(t)$, of the
temporal standard deviation of the mean filtered EEG to the mean
temporal standard deviation of the n EEGs, $R_{e}(t) =  SD_{T}$ of
$mean~ \underline{A}^{2}_{T}(t) / mean ~\underline{SD}_{T}$ of
$A^{2}_{j}(t)$, $j = 1,n$. When the oscillations were entirely
synchronized, $R_{e}(t) = 1$; when n EEGs were totally
desynchronized, $R_{e}(t)$ approached one over the square root of
the number of digitizing steps in the moving window.
Experimentally $R_{e}(t)$ rose rapidly within a few $ms$ after a
phase discontinuity indexed by $SD_{X}(t)$ and several $ms$ before
the onset of a marked increase in mean analytic amplitude,
$\underline{A}_{n}(t)$. That timing showed that the increase in
analytic amplitude could not be ascribed to synchronization, that
having already taken place. The succession of the high and low
values of $R_{e}(t)$ revealed episodic emergence and dissolution
of synchrony; therefore $R_{e}(t)$ was adopted as an index of
cortical {\it efficiency} (Haken, 1983), on the premise that
cortical transmission of spatial patterns was most
energy-efficient when the dendritic currents were most
synchronized. From this perspective the re-synchronization of
cortical oscillations after a re-initialization of their phases
was an energy-conserving step in a phase transition, preparatory
to a major increase in transmission intensity of a wave packet
after pattern stabilization.

These indices characterized four successive events comprising
cortical phase transitions forming wave packets. First
re-initialization of phase was marked by a rapid rise and drop in
$SD_{X}(t)$. Next came re-synchronization shown by a rise in
$R_{e}(t)$, followed by selection of an AM pattern, ${\bf
A}^{2}(t)$, and pattern stabilization indexed by $D_{e}(t)$.
Lastly the mean rate of free energy dissipation,
$\underline{A}^{2}(t)$, increased to a maximum in the time
interval before the next phase slip. Hence during a wave packet
cortex transmitted a synchronized carrier wave at high intensity
with stabilized AM pattern and frequency.

The times of onset of phase slips with respect to CS varied
unpredictably from trial to trial. Empirically (Freeman, 2005) we
found that the best predictor of the onset times of ordered AM
patterns was the ratio of the rate of free energy dissipation to
the rate of change in the order parameter, $H_{e}(t)$, because
$D_{e}(t)$ fell and $\underline{A}^{2}(t)$ rose with wave packet
evolution:
$$
H_{e}(t) = \frac{\underline{A}^{2}(t)}{D_{e}(t)} ~.
$$
We named this index {\it pragmatic information} after Atmanspacher
and Scheingraber (1990), and used it to construct tuning curves by
which to locate AM patterns and find the best settings for the
band pass filters needed for the Hilbert transform (Freeman,
2004b).

\section{Evidence for multiple overlapping wave packets in brain
states}

Application of the Hilbert transform to $64$ EEG signals in a
designated pass band gave an $8 \times 8$ matrix of analytic phase
values that were calculated at each digitizing step with respect
to the analytic phase from the spatial ensemble average of the
waveform at that step. Measurements were made at each step of the
spatial gradient of the phase in $rad/mm$ by fitting to the phase
surface a $2-D$ spatial basis function in the form of a cone
(Freeman and Barrie, 2000). The diameter at the base of the cone
was defined as the distance from the apex at which the phase
difference, $\Delta \phi$, from the phase at the apex was $\pi/4 ~
rad$, giving a soft boundary condition at half power where
$\cos^{2} \Delta \phi = 0.5$. The diameter served as a measure of
the prevailing extent of long-range correlation. The ratio of the
spatial gradient in $rad/mm$ to the temporal gradient (frequency
in $rad/ms$) gave the phase velocity in $m/s$, which gave values
that fell in the range of conduction velocities of axons running
parallel to the pial surface. The preservation of the phase
gradient and its conformance with conduction velocities implied
that the phase delay was imposed by the finite velocity of
communication in the cortex by axosynaptic transmission. Once
initiated, the oscillation in each local neighborhood was a
standing wave and not a traveling wave, which was crucial for
maintenance of stable spatial AM patterns. We inferred that the
sparseness of intracortical connectivity (Braitenberg and Sch\"uz,
1991) could not support strong entrainment of the oscillations,
thereby preserving the phase gradient for the $3-5$ cycles at the
center frequency of the ensuing oscillation in ordered AM
patterns. The location of the conic apex on the pial surface was
interpreted as the site of nucleation for the phase transition by
which an accompanying AM pattern formed. The sign of phase at the
apex was either maximal lead or maximal lag. Within frames the
sign was fixed and the changes in location were less than the
interelectrode interval that determined spatial resolution.
Between frames the location and sign changed randomly over
distances that often exceeded the dimensions of the square array.
Thus the analytic phase gave estimates of the diameters,
durations, locations in time and space, and the time required for
re-initialization in wave packets, but the phase gave no
information about behaviorally related content. That came solely
from calculations of the analytic amplitude from the Hlbert
transform (Freeman, 2004a).

Every AM pattern was accompanied by a conic phase pattern that
retained the history of its site of nucleation and spread, but
phase cones were also found between ordered frames and overlapping
with them at near and far frequencies. Our effort to determine the
mean and range of values for the parameters of phase cones led to
the discovery that the values had power-law distributions.
Moreover, the means and $SD$s of the distributions varied with the
size of the time window for calculating the rates of change of
phase in time and with the interelectrode distance in space
(Freeman, 2004b). Overlaps of $4$ to $6$ identified phase cones
were common, so that the phase cones in multiple EEG signals gave
the image of neocortex resembling a pan of boiling water with its
temperature kept at the boiling point by the release of energy in
bubbles, or fog with vapor continually condensing in droplets and
evaporating, or a sand pile kept at its critical angle by repeated
avalanches having power-law distributions of their sizes and
durations (Bak,1996).

As shown by Prigogine (1980) and Haken (1983) the approach of a
system far from equilibrium to a critical phase transition would
be manifested by slowing of frequency and of the rate of change in
the order parameter, and by increased amplitude of oscillations in
the output. These changes were revealed by $R_{e}(t)$, $D_{e}(t)$,
and $\underline{A}(t)$. We inferred that the emergence of order in
each cone was a spontaneous phase transition induced by the steady
input, implying that the neocortex held itself in a
pseudo-equilibrium state of self-organized criticality (Bak, 1996;
Jensen, 1998; Linkenkaer Hansen et al., 2001), so that the order
parameter, ${\bf A}^{2}(t)$, might vary with time, which it could
not do in a system at true equilibrium. We inferred that the
critical variable being stabilized in neocortex was the local mean
firing rates in neighborhoods of cortical neuropil engaged in
mutual excitation, which were homeostatically regulated everywhere
by the refractory periods of the neurons without need for
inhibition or any global monitor and feedback (Freeman, 1975/2004,
2000). Stable mutual excitation achieved a ground state of minimal
energy that provided the unpatterned background activity neurons
require in order to survive, the excitatory bias required for
oscillatory activity in the EEG by feedback inhibition, and a
state of readiness to jump to a different ground state under the
impact of weak inputs in accordance with constraints posed by
processes of intention and attention.

We came to view the background fluctuations in the EEG as
manifestations of its trajectories of continual relaxation toward
an optimal energy-efficient rest state of the cortex under
bombardment within itself and from subcortical modules including
the thalamus (Steriade, 1997) and basal ganglia (Houk, 2001). In
accord with Prigogine's conception the great majority of
fluctuations were quenched, but some crossed a threshold and
carried large areas of cortex into new domains of brain state
space (observed in n-space) that were expressed by new AM
patterns. The importance of sensory input for destabilization of
neuropil was obvious for the olfactory EEG, because AM patterns
formed with inhalations. Comparable roles of sensory inputs for
visual, auditory, and somesthetic cortices were inferred from the
onsets of repetitive ordered AM patterns within a few $ms$ after
onsets of CS and before the onsets of conditioned responses.
Moreover, during deep slow-wave sleep the AM patterns disappeared,
and occasionally the phase cones likewise disappeared, indicating
that EEG disclosed a basal state in sleep when sensory and other
inputs to cortex were blocked or withdrawn (Freeman, Holmes, West
and Vanhatalo, 2005) though not totally as in surgical
deafferentation (Burns, 1958; Becker and Freeman, 1968; Gray and
Skinner, 1988).

We observed the transition from awake rest to deep sleep in an
epileptic subject as graded without basic change in form until the
onset of intermittent episodes of a flat field and its equally
abrupt termination about $1~ s$ later. Whatever might be its
significance for the physiology of sleep, which is unknown, or for
epileptogenesis equally unknown, from the standpoint of the
application of field-theoretic models to nonlinear brain dynamics
this event offered an anchor for the theory in three respects.
First, the onset and offset for each flat epoch provided clear
instances of phase transitions. Second, the absence of spatial
patterns of both phase and amplitude constituted a high degree of
symmetry in the transient state during deep sleep. This state
approached that seen when EEG was flattened by deep anesthesia or
by surgical isolation of cortical slices and slabs. We described
this state of minimal spatiotemporal structure as a "vacuum state"
with minimal order parameters, to which we assigned the value of
zero. It provided a reference level for other states, most
immediately for the embedding state of sleep in which there were
phase cones but no accompanying AM patterns. Third, we interpret
the emergence of wave packets and the return of waking behaviors
as SBS to a more ordered regime characterized by a range of
parameters allowing repeated phase transitions to any of a range
of domains in brain state space having attentive engagement of the
subject with the environment with high likelihood of intentional
action.

In this state our animals were open to and searching for sensory
stimulation, operationally defined as CS. The receipt of expected
input precipitated phase transitions in which further symmetry
breaking took the form of an AM pattern (Freeman, 2005). The AM
patterns lacked invariance with respect to the CS under changes in
context and reinforcement contingency, so they could not be
representations of the CS.  Owing to dependence on context, we
inferred that AM patterns were shaped by the connectivity of
cortical neurons that had been formed by prior learning, which
formed an attractor landscape with basins of attraction
corresponding to the generalization gradients for categories of CS
that a subject had learned to discriminate (Ohl, Scheich and
Freeman, 2001; 2005).

We propose that a CS {\it selects} a basin of attraction in the
primary sensory cortex to which it is directed, often with very
little information as in weak scents, faint clicks, and weak
flashes. The astonishingly low requirements for information in
high-level perception have been amply demonstrated by recent
accomplishments in sensory substitution (Cohen et al., 1997; Von
Melchner, Pallas and Sur, 2000; Bach-y-Rita, 2004, 2005). There is
an indefinite number of such basins in each sensory cortical area
forming a pliable and adaptive attractor landscape. Each attractor
can be selected by a stimulus that is an instance of the category
that the attractor implements by its AM pattern. In this view the
waking state consists of a collection of potential states, any one
of which but only one at a time can be realized through a phase
transition. The variety of these highly textured, latent AM
patterns, their exceedingly large diameters in comparison to the
small sizes of the component neurons, the long ranges of
correlation despite the conduction delays among them, and the
extraordinarily rapid temporal sequence in the neocortical phase
transitions by which they are selected, provide the principal
justification for exploring the interpretation of nonlinear brain
dynamics in terms of many-body theory and multiple ground states
to complement basin-attractor theory.

\section{The spontaneous breakdown of symmetry in many-body
physics}

In this and in the following Section we switch to the qualitative
presentation of basic concepts in the many-body field theory with
spontaneous breakdown of symmetry. Our presentation is necessarily
focused solely on the few aspects that are most relevant to our
task. The interested reader may find full mathematical details in
Bratteli and Robinson (1979), and Umezawa (1993). For an extended
qualitative presentation, see Vitiello (2001). Our main goal here
is to illustrate how many-body theory bridges between microscopic
quantum dynamics and macroscopic behavior. The key point is the
existence implied by SBS of long-range correlation modes that are
responsible for ordering the system and therefore for the
collective behavior of the system components, which manifests in
the order parameter of classical fields characterizing the
macroscopic system behavior.

\vspace{0.3cm}

{\bf  4a. From microscopic quantum dynamics to the macroscopic

~~~~~behavior of the system}

\vspace{0.3cm}

Statistical mechanics is devised to deal with systems formed by a
large collection of indistinguishable components, such as a fluid
or a solid. The mathematical formalism allows statistical averages
and probability distributions (Ingber, 1995; Friston, 2000). An
early example of an application to cortical dynamics was by Wilson
and Cowan (1973), who used KII topology (Freeman, 1967) but
incorrectly opted for solutions as travelling waves not conforming
to the standing waves observed in wave packets. There are,
however, regularities observable at the macroscopic level such as
AM patterns that cannot be treated as regularities {\it only in
the average}, as the statistical treatment requires. The
complexity of patterning and the reproducibility with which such
regularities recur suggest that the AM patterns are generated by
dynamical processes, rather than by the random kinematics that
rule the collisions of the many elementary components on which the
formalism of statistical mechanics is based. These regularities,
such as the mesoscopic AM patterns in wave packets, typically
occur in quantum many-body systems. Examples of such ordered
patterns in physical systems are the spatial ordering of atoms in
a crystal or in a protein, the coherence in the time ordering of
phase oscillations of the photons in the laser beam, the coherence
in space ordering and in phase oscillation of the atomic or
electronic magnetic moments in the magnets, the strict temporal
sequences of interlocked chemical reactions such as the chains of
time ordered chemical steps in some metabolic activities (e.g.,
Davia, 2005). Dealing with realistic many-body systems means to
deal with an exceedingly high number of elementary components; it
is then necessary to describe the quantities of interest in terms
of fields.

The questions then arise: how to derive from the microscopic
quantum dynamics the regularities and the ordered patterns
observed at macroscopic level, and how to describe the process of
formation of these ordered patterns, namely the transitions from
the disordered phase, where no distinguishable patterns exist, to
the ordered phase? This involves a problem of {\it change of
scale}: from the microscopic scale (typically fixed by the Planck
unit of action $h$), where the dynamical description involves
quantum particles, to the macroscopic scale, where the system
macroscopic properties manifestly do not depend anymore on the
individual behavior of the particles, but on their collective
motion and are described in terms of classical fields, namely
collective waves or modes. At the macroscopic level, the relevant
scale for the resulting field theory is conditioned by the large
(as compared to the component typical size) volume scale. The
answers to such questions, widely tested in experiments, are
provided by many-body theory. There one can show that by SBS the
quantum field dynamics ruling the interactions of elementary
components predicts their collective motion (ordering) in the form
of macroscopic waves that characterize the system's classical
behavior (see e.g. Umezawa, 1993 and references therein).

\vspace{0.3cm}

{\bf 4b. Spontaneous breakdown of symmetry and phase transitions}

\vspace{0.3cm}

Let us recall that in the field theory of many-body systems the
fields may undergo symmetry transformations that leave the field
equations unchanged in their form. However, we are interested in
those states of the system that are {\it not symmetric} under the
symmetry transformations of the field equations. Indeed, there are
physical properties characterizing the physical behavior of the
system, which change during its evolution under certain boundary
conditions. When this happens we say that the system undergoes a
phase transition. As already mentioned, examples of phase
transitions are the ones leading to the formation (or, vice-versa,
to the destruction) of ordered patterns, e.g. the formation of a
crystal out of a solution of ions, the formation of a magnet out
of a collection of atoms carrying magnetic moment; the
condensation or evaporation of water; or the reverse of such
processes. As we will see, the possibility of formation of ordered
patterns is achieved when the constraints imposing the invariance
of the state of the system under symmetry transformations are
removed, so that the symmetry is broken. Order thus appears as the
loss of symmetry, and conversely symmetry is restored in the
transition from order to disorder. This conception may seem
paradoxical to neurobiologists, for whom bodies with mirror
symmetry are more ordered than those that are disfigured; however,
broken symmetry gives more information and hence more order.

One possible way to break the symmetry is to modify the dynamical
equations by adding one or more terms that are explicitly not
consistent with the symmetry transformations (are not symmetric
terms). This is called {\it explicit breakdown of symmetry}. The
system is forced by the external action into a specific
non-symmetric state that is determined by the designated imposed
breaking term. This is what we observe in the response of cortex
to perturbation by an impulse, such as an electric shock, sensory
click, flash, or touch: the evoked or {\it event-related
potential} (ERP). The explicit breakdown in cortical dynamics is
observed by resort to stimulus-locked averaging across multiple
presentations in order to remove or attenuate the background
activity, so as to demonstrate that the location, intensity and
detailed configuration of the ERP is predominantly determined by
the stimulus, so the ERP can be used as evidence for information
processing by the cortex.

The alternative possibility is to break the symmetry by submitting
the system to a weak, sustained stimulus: in cortex either by
prolonging a CS several tens of $ms$ or by relying on the
subcortical after-discharge from sensory impulse driving to
provide the prolongation. In such circumstances, in order to break
the symmetry, some pre-stimulus parameters, such as the coupling
constants (synaptic strengths) among the components or other
physically relevant parameters entering the dynamical equations,
must have values in well defined ranges, in which ranges we can
show by perturbation that the system is stable. These parameters
may depend on temperature and therefore their range of stability
in poikilotherms may be probed by varying the brain temperature.
The temperature of the mammalian brain is homeostatically
regulated, so instead we manipulate chemical potentials such as
the concentration gradients of extracellular ions,
neurotransmitters and neuromodulators, or we impose sustained
direct electric current through the cortical slices. One peculiar
property of quantum field dynamics, which makes it so successful
in the description of many-body systems presenting different
phases, and which motivates us to apply it to brain dynamics, is
that there are many of these stability parameter ranges, each one
characterizing a specific phase of the system with specific
physical properties that differ from phase to phase, or from each
AM pattern to the next. Each of the system phases is indexed by a
value of the macroscopic classic field which is called the order
parameter, ${\bf A}^{2}(t)$. If the dynamical regime is
characterized by that range of parameter values which does not
allow the breakdown of the symmetry, the system does not
perceptibly or meaningfully react (as in sleep to weak stimuli).
When one or more control parameters, such as the strength of
action at one class of synapse in the cortical pool under the
influence of the weak external stimulus, or even by indeterminate
drift, exceed the range of stability where the system originally
sits, the transition is induced to another stability parameter
range, different from the previous one in that it now allows
symmetry breakdown and appearance of order (as in arousal from
deep sleep). Contrariwise, the loss of order as in shutting down
under anesthesia or in deep sleep corresponds to symmetry recovery
or restoration, the formlessness of background activity or in the
extreme the loss of activity approaching brain death.

The symmetry breakdown is thus, itself, a property of the system
inner dynamics, dependent on the internal boundary conditions for
what concerns the appropriate control or tuning of the parameters,
but it is independent of the specificities of the triggering
stimulus, which may fall on differing equivalent sensory receptors
on successive trials. The stimulus characterizes the cortical
response in the explicit breakdown as in ERPs, irrespective of the
state of the system. In SBS the response of the system to the
breaking stimulus depends on the situation as it impacts the state
of the system, bringing in such biological and psychological
factors as prior learning, context, motivation, fatigue, etc. The
non-symmetric state into which the system transits (at the end of
the phase transition process) can be any one among those {\it
compatible} with the actual dynamical regime. The dynamics,
provided the system is in a parameter regime neighboring to the
previous range, constitutes a choice among the compatible
non-symmetric states available to the system, which in Section 3
we characterized as an attractor landscape. The breakdown of the
symmetry is then said to be {\it spontaneous}, in much the same
sense as the term is used to describe the background EEG which is
commonly labelled 'spontaneous'. Most obviously in experimentation
the time of onset of the SBS varies with respect to stimulus
onset, in contrast to the fixed latencies of explicit symmetry
breaking, and the emergent pattern is determined by the parameters
of the system, only remotely by the stimulus that has selected the
pattern. We say that the attractor governing the AM pattern is
{\it selected} and then modified by the stimulus.

\section{Order, symmetry and long-range correlations}

We have seen that the SBS is a microscopic dynamical phenomenon
that manifests in the generation of ordered patterns at the
macroscopic level; it involves a change of scale: from microscopic
to macroscopic. It correlates elementary components in an ordered
pattern. This is normally expressed also by saying that order is a
collective phenomenon and that SBS amounts to creation of a
communication, or correlation, or coherence among the elementary
components that extends over the system or over the domain of the
ordered pattern, a correlation of long range indeed. It might be
spatial coherence, as for spatial ordering, or temporal coherence,
as for in-phase oscillation, or both, as in cortex. The non-zero
value of the order parameter is a measure of the nature and degree
of coherence among the elementary system components. Our choice of
a vector as our order parameter is dictated by the complexity of
ordered brain states, such that the behavioral order cannot be
ranked by a scalar. Many-body communication, in order to be
effective, even at the lowest value of the order parameter, must
be robust and immune to distortion or failure. Therefore it cannot
originate only by short-range interactions. In such a case,
indeed, local defects or fluctuations would preclude the ordering
on a macroscopic scale, contrary to what it is observed in
many-body physics: for example, a dislocation in a crystal does
not destroy the whole crystalline ordering but deforms it only
locally, and a microlesion in cortex does not abolish large-scale
EEG or behavioral patterns.

Theoretical derivations and experimental observations in physical
systems show that SBS is always accompanied by the dynamical
formation of collective waves, the Nambu-Goldstone (NG) modes or
bosons or waves that span the whole system and sustain the
ordering communication. Physicists also say that these ordering
waves (or bosons) condense in the system ground state, and that
ordering is the result of a boson condensation. The phonons or
elastic waves detected in crystals and the magnons or spin waves
observed in magnets are examples familiar to physicists of these
carrier NG waves responsible for the respective macroscopically
observed ordered patterns. The modulus square of the amplitude of
these ordering waves for each given value of the associated wave
number (spatial wavelength) is proportional to their density (how
densely NG waves are present) in a given state of the system. Such
a density is thus a measure of the NG mode condensation: the
higher is the condensation, the higher is the degree of ordering.
Increasing (condensation) or decreasing (evaporation) the NG wave
density thus drives the system through its different physical
phases. In this way the boson condensation (or the evaporation
process), and therefore ordering (or disordering) process, becomes
possible due to the SBS mechanism (De Concini and Vitiello, 1976).
We propose that our ratio of variances, $R_{e}(t)$, can serve as
an experimental index of the density of condensation, and that our
Euclidean distance, $D_{e}(t)$, can serve as an index of the
stability of the condensate in cortical wave packets.

\section{Multiple ground states, external stimuli and
dissipation}

\vspace{0.3cm}

~~~~~{\bf 6a. Dissipation and a multiplicity of ground states}

\vspace{0.3cm}

We have seen that in the process of phase transition the system
evolves from one state to another, each with given physical
properties, for example, from a state with zero ordering to
another state with a non-zero value of the ordering, as from the
flat field of deep sleep to the background state of sleep and
beyond to a multiplicity of ground states. The existence of the
multiplicity characterized or, say, labeled by all possible values
of the vectorial order parameter is ensured by the mathematical
structure of the many-body field theory (Bratteli and Robinson,
1979). The possibility for the system to be in a state that is a
collection (or superposition) of co-existing and overlapping
ground states labeled by different values of the order parameter,
yet without, or with reduced, reciprocal interferences, results
from the mathematical structure of the field theory for
dissipative, open systems (Celeghini et al., 1992; Vitiello, 1995;
2001). These are systems that exchange energy with the environment
in which they are embedded. The brain is an example of an
intrinsically open system, in permanent although discretionary
interaction with the environment.

The possibility of the existence of a multiplicity of ground
states in dissipative systems can be understood in the following
way (see Appendix B, Part 2). Let us consider the situation in
which the system coupled with the environment evolves in time
through a sequence of states where the balance of the energy
fluxes at the system-environment interface including heat
exchanges, is reached, say $E = E_{Syst} - E_{Env} = 0$, with
$E_{Syst}$ denoting the system energy and $E_{Env}$ the
environment energy. Although the brain holds itself far from
equilibrium, this energy balance is manifested, for example, in
the homeothermic regulation of mammalian brain temperature and
body weight.  We call this equilibrium state the ground state; by
definition it is the $E = 0$ state. Clearly, balancing $E_{Syst} -
E_{Env}$ to be zero, does not fix the value of either $E_{Syst}$
or $E_{Env}$. It only fixes their difference. To each couple of
values $(E_{Syst},E_{Env})$ such that $E_{Syst} - E_{Env} = 0$,
there is one associated ground state. The important result
(Celeghini et al., 1992; Vitiello, 1995) is that each couplet of
specific values $(E_{Syst},E_{Env})$ uniquely characterizes an
associated ground state; in different words, ground states
corresponding to couplets of different values $(E_{Syst},E_{Env})$
may overlap without (or almost without) destructive interference.
We may then use the value of $E_{Syst}$ as a label to specify the
ground states in the brain at any given instant of time. Moreover,
we can show that what makes the difference among ground states of
different label values is their different degree of ordering due
to some collective mode. These states denote the physical phases
of the system. The relations thus found among the energy
$E_{Syst}$ include heat, $Q$, free energy and its rate of
dissipation estimated by $\underline{A}^{2}(t)$, and our vectorial
order parameter that measures the ordering, ${\bf A}^{2}(t)$,
labelled by the scalar $\underline{A}^{2}(t)$ (Section 2). Then
$E_{Syst}$ and its variations are directly related to the
collective mode density (Appendix B, Part 2 and 4), which is
indexed by the boson condensation density, $R_{e}(t)$. As we have
already observed, ordering means spatial structure seen in the
emergence of phase cones and of spatial patterns measured by ${\bf
A}^{2}(t)$ and $D_{e}(t)$, as well as temporal synchrony measured
by the variance of the instantaneous phase differences across the
array, $SD_{X}(t)$, and by the ratio of variances, $R_{e}(t)$
(Section 2). No one of these parameters suffices to capture the
relations in a single scale.

The brain in its relation with the environment, owing to its
property of being a dissipative or open system, may occupy any one
of this multiplicity of ground states, depending on how the $E =
0$ balance or equilibrium is approached. Or else, it may sit in
any state that is a collection or superposition of these
brain-environment equilibrium ground states. The system may shift,
under the influence of one or more stimuli acting as a control
parameter, from ground state to ground state in this collection
(from phase to phase) (Appendix B, Part 2, 4 and 5), namely it may
undergo an extremely rich sequence of phase transitions, leading
to the actualization of a sequence of dissipative structures
(Prigogine, 1980) formed by wave packets.   The system {\it
trajectory} through these phases may also be chaotic (Pessa and
Vitiello, 2003; 2004) (Appendix B, Part 5) and {\it itinerant}
through a chain of 'attractor ruins' (Tsuda, 2001) constituting a
set of attractor landscapes (Skarda and Freeman, 1987) accessed
serially or merely approached in the coordinated dynamics of a
metastable state (Kelso, 1995; Bressler and Kelso, 2001;
Fingelkurts and Fingelkurts, 2001).  We stress that the
possibility of deriving from the microscopic dynamics the
classicality of such trajectories is one of the merits of the
many-body field model. For example, we know that multiple wave
packets having different carrier frequencies co-exist in each
sensory cortex, and that one or more of these is local while one
or more is global, yet we do not yet know whether or how these
wave packets might influence each other in the determination of
sensory cortical output. Many-body physics may provide the tools
needed to measure, analyze and explain these complex events.

\vspace{0.3cm}

{\bf 6b. Selection of AM patterns by conditioned stimuli}

\vspace{0.3cm}

In Section 3 we saw that the cortical standing wave resulting from
a phase transition forming a wave packet was given the appearance
of a traveling wave by the delay in initialization embodied in the
phase cone. On successive trials with the same CS the location of
the apex varied randomly within the primary receiving area for the
CS modality, and its sign (maximal lead as in an explosion or
maximal lag as in an implosion) likewise varied randomly from each
transition to the next. These random variations gave further
evidence for SBS. The sudden change in system with order parameter
evolution could be classically described in terms of a subcritical
Hopf bifurcation or departure from an unstable point repellor at a
saddle node. Indeed, this would agree with the theoretical feature
that variations in an order parameter can serve to describe
classical trajectories in the order parameter space, which appear
to be chaotic (Appendix B, Part 5) (Pessa and Vitiello, 2003;
2004; Vitiello, 2004).

At first view the AM patterns appeared to be cortical
representations of CS that differed in the same way that activity
patterns in the retina, skin, nose or cochlea differed on repeated
presentation of the same CS. However, the patterns that were
elicited by an invariant CS held only within each training session
and then only if there were no changes in the schedule of
reinforcement or addition of a new CS in serial conditioning.
Measurements of AM patterns within sessions showed pattern
variation within each category despite CS invariance. Between
sessions with no new CS added the averages of the patterns tended
to drift. When the subjects were trained to respond to a new CS,
or when the reinforcement was reversed between CS+ and CS-, all of
the patterns changed including the pattern for the background, and
the amount of change with new learning was 2 to 4 times the
average change with drift across multiple sessions. A collection
of AM patterns that we established by training persisted with
drift through multiple sessions until we introduced the next
contextual change (Freeman and Grajski, 1987). Pattern variation
without change in CS but with change in context and import showed
that the AM patterns were determined mainly by the cortical
dynamics, as the many-body model predicted, and only marginally by
the incoming stimulus. This inference was consistent with the fact
that most neocortical synapses on the pyramidal cells generating
the EEG came from neurons in the cortex; only a small fraction
came from thalamic neurons.

Here we describe the engaged state as based in a collection of
ground states that differ from the awake rest state in being
potentially distinguishable from each other, if and when they are
accessed by a relevant input. The set of these ground states
constitutes a family of {\it compatible} states. The possibility
to access any one of such a collection of states is evidence of
the dissipative character of the many-body dynamics, as noted in
Section 6a. Evidence for dissipation is that the ground states
during engagement, when actualized in the post-stimulus period,
differ in having levels of power above or below that of the
undifferentiated pre-stimulus control state but always above the
awake rest state. The dissipative many-body model predicts that
these power differences relate to energy differences caused by the
boson condensation process, owing to variations in the density and
structure of the condensate, as discussed in Section 9 (see also
Sections 5 and 6 and Appendix B, Part 4).

\vspace{0.3cm}

{\bf  6c. Levels of activity achieved by SBS or its reverse}

\vspace{0.3cm}

Summarizing, we have identified five levels of cortical activity
that we describe by the dissipative many-body field theory model:

a) The transient "vacuum" state of the cortex is characterized by
unbroken symmetry: a flat field of fluctuations with no
discernible patterns. The system dynamical regime is characterized
by parameters whose values do not allow for the symmetry
breakdown. Even in the presence of an external stimulus (provided
it is below a threshold) the system "cannot" react to it
(breakdown is not possible). The system "sleeps". External inputs
at most create uninteresting perturbations.

b) These transient epochs of deep sleep interrupt a longer-lasting
state of slow-wave sleep characterized by multiple short-lasting
and overlapping phase cones. We suggest that these fluctuations
result from continuous bombardment of all areas of neocortex by
other parts of the brain, including inputs from the sensory
receptors that are relayed mainly through the thalamus and that
are mainly irrelevant, because it is the work of cortex by
habituation to establish filters to mitigate the impact of such
unavoidable bombardment on cortices by irrelevant energies from
the environment. The continual perturbation gives rise to myriad
local phase transitions characterized by the conic phase
gradients, which are quenched as rapidly as they are formed,
thereby maintaining the entire cortex in a robust state of
conditional stability (called metastable by Kelso (1995), Bressler
and Kelso (2001) and Fingelkurts and Fingelkurts (2004)) that we
propose conforms with self-organized criticality (Bak, 1996;
Jensen, 1998; Linkenkaer-Hansen et al., 2001). The critical
parameter is the mean firing rate of neurons that is
homeostatically maintained by mutual excitation everywhere by
thresholds and refractory periods (Freeman, 2004b). The phase
cones are brief with no distinguishable AM patterns, so we infer
that they are related to SBS with vanishingly short-lived order
parameters.

c) In the awake rest state the ranges of parameter allowing
higher-order SBS (cf. Section 4) become potentially accessible
under the influence of external weak but behaviorally significant
stimuli. The temporal phase differences appear as coordinated
analytic phase differences (CAPD, Freeman, Burke and Holmes, 2003)
in which each epoch of minimal $SD_{X}(t)$ is accompanied by a
peak in mean amplitude, but without discernible or reproducible
spatial AM patterns, owing to the lack of engagement of the
subject with the environment.

d) The rest state evolves into an aroused state with increased
amplitude of oscillations in the background dendritic current that
accompanies incipient engagement of the brain with the external
world through the body. There is an implicit differentiation among
the set of compatible states, which is only realized by the overt
emergence of an amplitude pattern as reported above that is
classifiable, and that arises from SBS triggered by a relevant
stimulus that breaks the symmetry of the expectant state. The
sequence of states is reversed on the return from engagement to
expectancy through rest to sleep and then to transients with
maximal disorder.

e) A departure from this sequence was observed (Freeman et al.,
2005) prior to the onset of a complex partial seizure consisting
of spikes at $3/s$ accompanied by loss of consciousness
("absence") and stereotypic motor automatisms. During the
pre-ictal phase a substantial reduction was calculated in the
diameters of phase cones, owing to doubling of the phase gradients
and implying a significant impairment of long-range correlation,
preceding the loss of spatially coherent carrier waves. This
observation attests to the importance of long-range correlation
for the maintenance of normal metastability of the cerebral
cortex. The seizure was simulated (Freeman, 1986; Skarda and
Freeman, 1987) with solutions of differential equations
constituting a model of the dynamics of the olfactory system
(Shimoide and Freeman, 1995; Freeman, 1998; Li, Li et al. 2005;
Li, Lou et al 2005). We use the model to infer that the loss of
long-range correlation prior to seizure onset is due to abnormal
levels of activity of inhibitory neurons from unspecified causes,
and that the loss releases local domains from cooperative activity
and enhances spatial contrast by lateral inhibition, hence an
instability by mutual inhibition constituting a form of positive
feedback.

\section{Amplitude and phase patterns suggest a carrier for
long-range communication}

A detailed examination of the wave packets in the visual, auditory
and somesthetic cortices in rabbits and cats under a classical
aversive conditioning paradigm has shown that between the arrival
of a CS and the execution of a conditioned response (CR) there
were typically $3$ to $4$ ordered AM patterns with their attendant
phase cones (Freeman, 2005). The properties of the first and
second (on some trials) wave packets had the following properties:
carrier waves in the gamma range ($30-80 ~Hz$), durations seldom
exceeding $100 ~ms$; diameters seldom exceeding $15 ~mm$; low
power in the $1/f^{a}$ relation; repetition at rates in the high
theta range ($5-7 ~Hz$); and relatively steep phase gradients
corresponding to lower phase velocities. The later wave packets
had carrier frequencies in the beta range ($12-30 ~Hz$); durations
often exceeding $100 ~ms$; estimated diameters large enough to
include multiple primary sensory areas and the limbic system;
greater power by $1/f^{a}$ (as predicted by the model, see above);
repetition rates in the low theta range ($3-5 ~Hz$); and shallower
phase gradient with, in some cases, extreme phase velocity beyond
measurement. Whereas the gamma events were found infrequently in
the pre-stimulus control period, the beta events appeared with
equal frequency in the pre-stimulus control and post-stimulus test
periods.

Both types of wave packet had onsets by phase transition based in
SBS but with differing forms or degrees of symmetry. The beta and
gamma carrier waves were clearly macroscopic properties of the
populations of cortical neurons that were responsible for the
oscillations, and the waves were clearly manifested not in the
individual pulse trains of the neurons by time-locked firing but
by oscillations in the probability or relative frequency of firing
of the neurons in the domain, especially when, as was typical, the
neurons only fired pulses at time intervals on average much longer
than the mean wavelengths of the oscillation in $ms$ (Freeman,
1975/2004). The formation and maintenance of the shared
oscillations with inconstant instantaneous frequencies, $\omega
(t)$, depended on rapid communication among the neurons
effectively out to the spatial soft boundary condition (Section 3)
in the two surface dimensions of the cortical domain of coherence.
That communication required a channel to convey influence from
each part of the domain to every other part. Reasons have already
been given in Section 1 for the inadequacy of classical
explanations based on nonsynaptic diffusion, the action potential,
Coulomb forces, or electric, magnetic and radio fields. The
difference between the slow rate of serial synaptic transmission
of energy laterally in cortical networks and the high phase
velocity of phase cone formation was described as anomalous
dispersion without further explanation. That concept may be
inadequate to encompass beta and gamma phase transitions. The
field theoretic model offers an alternative view of the phase
transition as a condensation comparable to the formation of fog
and raindrops from water vapor that might serve to model both the
gamma and beta phase transitions.

That leaves the resolution of the problem of long-range
correlation and the genesis of macroscopic order parameter fields
to the mechanism of SBS in the dissipative many-body model. The
boson condensate that emerges within the domain of the phase cone
has one AM pattern from a compatible set of distinguishable ground
states that is latent in the synaptic connectivity of the domain
comprising the collection of sub-areas making up each primary
receiving area with synaptic connections that have been shaped by
prior learning. The rapid re-initialization is then enabled
through the relatively sparse long axons linking all cortical
areas that constitute an essential link in the chain between
atomic and molar phenomena. Long axons require large cell bodies,
complex mechanisms for guidance during brain development, and
expensive blood supplies for maintenance and operations to provide
the fast communication that enables small-world effects. Their
action potentials shape the properties of the spatial and temporal
patterns described above at the mesoscopic and macroscopic levels
that cannot be derived only from properties at the atomic and
molecular levels. In the absence of these SBS waves the action
potentials of long axons cannot easily explain, as already said
above, the spatial coherence often with negligible phase
dispersion at every transmission frequency.

These considerations raise the question of the physical nature of
the SBS carrier wave describing the diffuse synaptic influence. We
propose in answer that the carrier wave is a dipole wave in which
the $3-D$ rotational (electric) dipole symmetry is spontaneously
broken (Del Giudice et. al. 1985; 1986; 1988; Vitiello 1995; Jibu
and Yasue, 1992; 1995). The physical justification for this
inference is rooted in the fact that most biochemical molecules
sustaining the neural activity and the water matrix in which the
neural activity is embedded are physically characterized by
possessing non-zero electrical dipole moments (not to be confused
with cortical dipole fields engendered by sources and sinks of
dendritic current). The crucial point to be stressed again is that
the SBS mechanism involves the change of scale, from the
microscopic dynamics to the macroscopic order parameter field
dynamics. The system is thus a macroscopic quantum system, which
means that the observed classical behavior of the system cannot be
explained without resorting to the quantum dynamics of the SBS. It
is not the Planck constant scale that dominates the observed
phenomenology, but, rather, the long range of the observed
correlations in accord with the {\it large volume limit} that
dominates field theory.

We propose that the NG cortical boson condensate or {\it
symmetron} in the Ricciardi and Umezawa original terminology
(1967) might be used to explain the rapid course of perception:
how for example neocortex can respond selectively to the impact of
photons from a face in a crowd on a handful of retinal neurons
that transmit impulses mixed among uncountable impulses elicited
by light from the crowd. The further impact of the action
potentials on the visual cortex elicits countless more action
potentials relating to all the edges, points, colors, motions, and
contours in both background and foreground of the scene. Then
suddenly there emerges by phase transition order from disorder:
the neural "vapor", as it were, condenses into neural droplets,
the first step in face recognition within the time lapse from
stimulus arrival at the cortex of at most a few tens of $ms$,
which is insufficient to allow operations of gradient descent in
classical models.

\section{Superposition and the emergence of meaning}

The gamma condensate is only the first of two or more phase
transitions between the CS and perception leading to the CR, and
it appears to be restricted to the primary receiving area for each
modality. It can suffice to solve the "binding problem" (Singer
and Gray, 1995), but it does not support the multimodal
recognition that accompanies the sight of a face, the sounds of
voices, the press of bodies, and the odors of sweat and perfume,
in a word, the multisensory information inflow through all of the
sensory ports. The responses of the several receiving areas are
already shaped by attractor landscapes assisted through
preafference (Kay and Freeman, 1998; Kozma, Freeman and Erdí,
2003) to the expected sights, sounds, pressures and smells in the
situation, so that each area constructs and maintains a collection
of ground states {\it in potentia} as the basis for expectancy and
selective attention. The dissipative many-body model allows indeed
the possibility of classical chaotic trajectories in the
high-dimensional order parameter space, the space whose local
hypervolumes signify the ground states (Pessa and Vitiello, 2003,
2004; see Section 6). The arrival of an expected stimulus triggers
SBS, selects a ground state (cf. Appendix B), and places the
dynamics of the relevant cortex into the basin of an attractor,
which abstracts by deletion of nonessential information as the
trajectory spirals to the attractor, and generalizes to the
category of the stimulus that is reported by the cortex in the
output of an AM pattern. The several cortices transmit their
signals by broadcasting action potentials organized in gamma wave
packets to each other and to the limbic system. The information
carried by the wave packets is spatially distributed in the manner
of a hologram, so that every subset of axons from a cortical area
transmits the same information with the resolution proportional to
the size of the subset. The entorhinal cortex integrates all of
the transmitted signals; orientation in space and time are
introduced in the hippocampus; and a combined signal is broadcast
to all of the primary areas (Freeman, 2001; Kozma, Freeman and
Erd\'i, 2003), priming the several cortices for the formation by
SBS of a series of AM patterns that form carried by beta wave
packets. We infer that beta wave packets are multi-sensory events
manifesting priming of the primary sensory cortices by
preafference in anticipation of the arrival of a CS, and that they
update all of the cortices within the time required for the limbic
passage (Freeman and Burke, 2003; Freeman and Rogers, 2003). We
conclude that this cycle provides the dynamic structure of
multi-modal attention.

The concept of the boson carrier and the boson condensate does
more; it enables an orderly and inclusive description of the phase
transition that includes all levels of the macroscopic,
mesoscopic, and microscopic organization of the cerebral patterns
that mediate the integration of the animal with its environment,
down to and including the electric dipoles of all the myriad
proteins, amino acid transmitters, ions, and water molecules that
comprise the quantum system. This hierarchical system extending
from atoms to whole brain and outwardly into engagement of the
subject with the environment in the action-perception cycle is the
essential basis for the ontogenetic emergence and maintenance of
meaning through successful interaction and its knowledge base
within the brain. By repeated trial-and-error each brain
constructs within itself an understanding of its surround, which
constitutes its knowledge of its own world that we describe as its
double (Vitiello, 2001). It is an active mirror, because the
environment impacts onto the self independently as well as
reactively. The relations that the self and its surround construct
by their interactions constitute the meanings of the flows of
information that are exchanged during the interactions. We stress
once more that neuronal cells and other macroscopic structures are
by no means considered quantum objects in our analysis. No
ambiguity should be born on this point. The SBS provides in the
ways above explained the bridge, or change of scale, from the
microscopic quantum dynamics to the macroscopic behavior of
classical cells and their constructs.

\section{Free energy, dissipation, ordering, and fMRI}

In this section we present further details of our analysis by
interrelating free energy, ordering and stability of the amplitude
patterns. Neural activity exists in pulse densities of axons, wave
densities of dendrites and in various forms of thermal, electric,
magnetic and especially chemical energy. At the mesoscopic and
macroscopic levels of neural populations the forms of neural
activity constitute a group invariant over the transformations
among these vehicles, subject only to the minimization of the loss
of free energy required to stabilize brain patterns in the overall
transit of energy carried to the brain by the arterial supply as
glucose and removed by the venous return as waste heat. The most
readily accessible measure of the rate of free energy dissipation
over the diverse conversion operations is the square of mean
analytic amplitude, $\underline{A}^{2}(t)$ (Section 3), in the
beta or gamma pass band.

The desirability of using $\underline{A}^{2}(t)$ as an index of
power and therefore indirectly of the rate of heat dissipation is
made possible by the sequence of four stages with each phase
transition. At the time of the fourth stage the phase transition
concludes with a dramatic increase in A2(t) to a peak value, at
which the EEGs remain synchronized and the AM pattern and carrier
frequency remain stabilized. At the peak the AM patterns are
maximally distinguishable between those on trials with a
reinforced CS+ and those without reinforcement CS- (Freeman,
2005a). During each frame a wide area of neocortex transmits $3-5$
cycles of synchronized wave carrying a stable AM pattern at high
intensity synaptic current controlling volleys of action
potentials. These volleys carrying the same signal on multiple
divergent pathways cannot fail to have significant impact on
neural networks and populations in the brainstem, cerebellum and
spinal cord, so the boson condensate in the frame provides a
plausible descriptor of the neural mechanism for neocortical
read-out by divergent-convergent axonal pathways that perform
spatial integral transformations on the output (Freeman, 2000).

The cost in dissipated free energy of the electrochemical work
done to create the AM pattern is high; the brain has only $2 \%$
of body mass, yet it consumes $20 \%$ of basal metabolic energy,
which is measured in heat, $Q$, and entropy, $S$, that must be
discarded to avoid increases in temperature and pressure in the
fixed volume and mass of the brain. These and other waste products
are removed by continuous adaptation of blood flow to rates of
energy dissipation. The poorly understood homeostatic link between
metabolic demand and arterial perfusion is the thermodynamic basis
for hemodynamic techniques of brain imaging that have been derived
from theoretical physics: fMRI, PET, SPECT and BOLD. What they
lack in temporal resolution is more than compensated by their
broad spatial views at high resolution, so that the
electroencephalographic and hemodynamic approaches are strongly
complementary. As noted earlier, the analytic amplitude squared is
the sum of squares of the real part proportional to the energy
expended in the current of excitatory neurons and the imaginary
part in quadrature proportional to the energy expended by
inhibitory neurons in negative feedback (Freeman, 2005b), so that
$\underline{A}^{2}(t)$ offers an optimal parameter of neural
activity for correlation with indirect measures of blood flow,
especially in the beta and gamma ranges of EEG oscillations. In
most physical systems the energy expenditure is proportional to
the frequency of oscillation; the fact that the observed power in
gamma waves is only $1 \%$ of that in theta waves in the $1/f^{2}$
clinical $PSD_{T}$ may be misleading. The actual metabolic costs
may be comparable in the two frequency ranges, so when properly
normalized the gamma power may be the more strongly correlated
with levels of metabolic demand.

\section{Conclusions and outlook}

We believe we have made a strong case to identify the EEG phase
discontinuities and the neocortical phase transitions with SBS,
and the multiplicity of coexisting phase cones and AM patterns
with the ground states predicted by the dissipative many-body
model. We have described our case in qualitative terms without
equations in order to make it intelligible to physicists,
neurobiologists, psychologists, engineers, and cognitive theorists
having diverse backgrounds and languages. We have emphasized that
our proposed synthesis depends heavily on new experimental data
derived from novel techniques for measuring brain field potentials
with unprecedented spatial and temporal resolution, and for
exploring dissipative dynamics in the context of quantum field
theory. We have cited the extensive mathematical structures on
which we base our descriptions of EEG data in Sections 2 and 3 and
of basic concepts in many-body theory in Sections 4 to 6, and we
have included Appendices that summarize our mathematics. Thirty
years ago we stepped beyond linear analysis into nonlinear
neurodynamics; now the new data are carrying us to new and
baffling levels of complexity.

Therefore in Section 7 we have introduced for further exploration
the identification of the wave packet with the Nambu-Goldstone
boson as a carrier for long-range correlation with two main
classes or types having the beta and gamma carrier waves. We
foresee substantial experimental and theoretical difficulties in
determining whether these bosons superimpose or interact, each by
modifying the other, and in extending the analysis into the
molecular realm in terms of theories such as those of volume
transmission of Fuxe and Agnati (1991) and Bach-y-Rita (2005) and
the catalytic model of Davia (2005). In Section 8 we have posed
the unsolved problems of extending many-body theory to modeling
the higher cognitive functions of the brain (e.g., Pribram, 1971)
and the construction of the knowledge base by which brains sustain
meaningful interactions with their environments. In Section 9 we
offer for exploration a link to current techniques in brain
imaging.

We propose that a promising area of research will be the found in
tapping the vast reservoir of information in scalp EEGs of human
subjects, who can report their experiences as their spatiotemporal
structures evolve, both normal and pathological. A prime
difficulty will be to adapt the theory flexibly to treat the
unique properties of neuropil and avoid forcing onto the models
the conditions that hold for other forms of condensed matter. It
is essential that neurobiologists and theoretical physicists
listen carefully to each other in order to determine what can and
should be done next.

\vspace{0.5cm}

\large{\bf Acknowledgements}

\vspace{0.3cm}

\normalsize

We are grateful to Harald Atmanspacher, Eliano Pessa and Terence
Barrett for reading the manuscript and for useful comments.

\newpage

\large{\bf Appendix A}

\vspace{0.5cm}

\normalsize

We describe the mesoscopic neural activity of neocortex as
comprising K-fields (activity density functions) having properties
corresponding with a hierarchy of neural groups we have designated
as K-sets (Freeman, 1975/2004), while the electric and magnetic
potentials by which we indirectly observe the K-fields are the
classical fields of Faraday and Maxwell. The most fundamental
physical description is of the topology of neocortex: a bounded
assembly of $10^{10}$ neurons (estimated in humans) and $10^{14}$
synapses forming functional connections that constitute a unified
field of neuropil in each hemisphere. The neuropil extends without
anatomical interruptions or obstructions to the lateral
transmission of neural activity to the boundaries of the neocortex
at the midline and entorhinal fissures. In humans the two
neocortices have a surface area approaching one square meter with
a thickness of $1$ to $3$ millimeters. Multi-electrode EEG signal
observations and measurements are made in the two surface
dimensions, so the dynamics is comparably developed in $2-D$, in
the sense that photographs and phonograph records are $2-D$.
Embedded within this neuropil are well-known specialized domains
of synaptic connectivity that are induced, elaborated, and
maintained by axonal projections from sensory systems and to motor
systems. The otherwise near-random local connectivity is sparse,
with each neuron connecting to only about $1\%$ of neurons within
the radius of its dendritic arbor (Braitenberg and Sch\"uz, 1991).
A small proportion of axons form connections that extend far from
nearest neighbor, up to the length of each hemisphere. These long
axons enable the distant transmission of activity that can create
seeds that spread locally, enabling the emergence of 'small-world'
effects (Watts and Strogatz, 1998; Kozma et al., 2005), so each
neuron is at most only $4$ or $5$ synapses distant from any other
in neocortex.

The neocortical neurons also form connections with neurons in
subcortical structures in the striatum, diencephalons, midbrain,
pons and (through relays) cerebellum that loop back in
multisynaptic chains. These connections (Houk, 2001) requiring
$3-D$ representation impose multisynaptic delays. These loops
together with the corticocortical long connections support
spatiotemporal patterns of neural activity that provide in each
cerebral hemisphere the organization of motor output (Houk, 2001)
and the predictions of the sensory input that result from acts of
observation. Histograms of the lengths of these connections form
approximate power-law distributions, suggesting the connectivity
is fractal (Wang and Chen, 2001) with a small number of sites
having very high numbers, which are sites of catastrophic loss of
function from very small lesions, such as those to the midbrain
reticular formation and the perforant path of the hippocampal
formation.

We elaborate on this basic topology of the neocortex with a
hierarchy of K-sets. It rests on the KO set used to model a $2-D$
array of neurons, either excitatory (KOe) or inhibitory (KOi),
with no connections among them but with input and output
connections for every neuron. The dendritic dynamics is modeled by
a linear second order ordinary differential equation (Freeman,
1967, 1975/2004) in the state variable, $v$, representing the
magnitude of neural activity in the wave density mode:
$$
\frac{d^{2}v}{dt^{2}} +a \frac{dv}{dt} + b v = I~,
$$
where $a$ is the rise time of a postsynaptic potential, $b$ is the
decay time, and $I$ is impulse input. The magnitude of neural
activity in the pulse density mode on axons, $p$, was modelled
(Freeman, 2000) with the solution to a second order static
nonlinear equation in normalized coordinates:
$$
        p = p_{o} \{1 + \exp[ - (e^{v} -1) / Q_{m} ]\},
$$
where $p_{o}$ is the mean background pulse density and $Q_{m}$ is
the maximum normalized pulse density. The KO set models the
dynamics of an average neuron that has only a zero point
attractor, to which it returns after perturbation. All interesting
activity patterns come from interactions among KO sets. The KIe
(or KIi) set models a collection of excitatory neurons that excite
each other by coupling two KOe (or KOi) sets in positive feedback.
If the density of connections exceeds a certain threshold, the KOe
set can sustain a non-zero point attractor. During development all
neurons at KO begin to sprout axons and dendrites extending widely
into their surrounds. They must repeatedly fire action potentials
in order to survive and at first do so by leaky membranes, but
when the threshold is reached when each neuron receives more
pulses back than it transmits, the activity of the now KIe
collection is sustained without need for membrane leakage.

In a $2-D$ KIe set a uniform field of random pulses and $1/f$
dendritic waves emerges that is limited by thresholds and
refractory periods without need for inhibitory neurons, which
emerge later in development. When they do so at around birth, a
KII set forms by the union of KOe and KOi sets. When sufficiently
dense negative feedback connections form, a limit cycle attractor
emerges with a characteristic frequency in the gamma range. The
KII sets model local collections of neurons in cortex with
conduction delays among them that are small enough to be absorbed
into the synaptic and dendritic delays of the rise time of a
postsynaptic potential (PSP) prior to passive membrane decay. The
dynamics of the component KO sets are modelled by a second order
ordinary differential equation performing integration on synaptic
input prior to the operation of bilateral saturation at the output
modelled with a static sigmoid function.

The assembly of three KII sets with long feedback connections
among them is modelled by a KIII set (Kozma and Freeman, 2001)
that gives a chaotic attractor landscape sufficing to model the
EEG and the operations of sensory cortices in categorizing
conditioned stimuli (CS) after learning (Freeman, 2000).

The KIV set serves to model the primitive forebrain by the
assembly of three KIII sets representing the sensory and motor
cortices and the hippocampal system that provides for short-term
memory and a cognitive map required for animal and robotic
navigation (Kozma, Freeman and Erd\'i, 2003).

The KV set models the dynamics of the neocortex in each cerebral
hemisphere. The undivided neocortical neuropil embeds the
ipsilateral primary sensory and motor areas with the limbic
system, and by its connectivity patterns provides the dynamics
that is required for the rapid global integration and evolution of
behavior. The observable mesoscopic electric fields are all
generated by KI sets, but their spatiotemporal patterns are
governed by their embedding in higher order K-sets together with
their controlling parameters.

\vspace{0.7cm}

\large{\bf Appendix B}

\vspace{0.5cm}

\normalsize

Some aspects of the mathematical formalism of the dissipative
brain model are here very shortly summarized for the reader
familiar with the terminology and the mathematics of QFT. For a
more detailed account of the formalism we refer to the original
papers (Vitiello, 1995; Alfinito and Vitiello, 2000, 2002;
Vitiello, 2004; Pessa and Vitiello, 2003, 2004).

\vspace{0.5cm}

{\bf {\it Part 1}}

\vspace{0.2cm}

As already mentioned in the text, patterns of correlated elements
(ordered patterns) which are macroscopically observable in
physical systems are described in QFT by the mechanism of
spontaneous breakdown of symmetry (SBS). Symmetry is said to be
spontaneously broken when the Lagrangian is invariant under
certain group of continuous symmetry, say $G$, and the vacuum or
ground state of the system is not invariant under $G$, but under
one of its subgroups, say $G'$. The ground state then exhibits
observable ordered patterns corresponding to the breakdown of $G$
into $G'$. These patterns are generated by the coherent
condensation in the ground state of massless quanta called
Nambu-Goldstone (NG) particles, or waves, or modes. These modes,
which are the carriers of the ordering information in the ground
state, are dynamically generated by the process of the breaking of
the symmetry. They manifest themselves as collective modes since
their propagation covers extended domains, or, in the infinite
volume limit, the whole system. The observable specifying the
degree of ordering of the vacuum (called the order parameter) acts
as a macroscopic variable for the system and is specific of the
kind of symmetry into play. Its value is related with the density
of condensed NG bosons in the vacuum. Such a value may thus be
considered to be the {\it code} specifying the vacuum of the
system (i.e. its macroscopically observable ordered state, namely
its physical phase) among many possible degenerate vacua.

In the quantum model of the brain (Ricciardi and Umezawa, 1967;
Stuart, Takahashi and Umezawa, 1978; 1979) the code of the ground
state specifies its memory content: the memory recording process
is depicted by the NG boson condensation in the brain ground
state. The external informational input acts as the trigger of the
symmetry breakdown out of which the NG bosons and their
condensation are generated. The symmetry which gets broken is the
rotational symmetry of the electrical dipoles of the water
molecules (Jibu and Yasue, 1995) and the NG modes are the
vibrational dipole wave quanta (DWQ) (Del Giudice, et al. 1985;
1986; 1988). The recall of the recorded information occurs under
the input of a stimulus ''similar'' to the one responsible for the
memory recording.

\vspace{1cm}

 {\it Part 2}

\vspace{0.3cm}

In the dissipative brain model a central role is played by the
fact that the brain is an {\it open system} continuously {\it
linked} (coupled) with the environment, namely that its dynamics
is intrinsically dissipative (Vitiello, 1995). The procedure of
the canonical quantization of a dissipative system requires  the
"doubling" of the degrees of freedom of the system (Celeghini,
Rasetti, Vitiello, 1992). Such a requirement ensures that the flow
of the energy exchanged between the system and the environment is
balanced.

Let $A_{k}$  and ${\tilde A}_{k}$ denote the annihilation
operators for the DWQ mode and its "doubled mode", respectively.
$k$ denotes the momentum and other specifications of the $A$
operators (similarly, we denote by $A^{\dag}_{k}$ and ${\tilde
A}^{\dag}_{k}$ the creation operators).

The initial value problem is defined by setting the {\it code}
${\cal N}$ imprinted in the vacuum at the initial time $t_{0}=0$
by the external input and representing the {\it memory record} of
the  input. The code  $\cal N$ is the set of the numbers ${\cal
N}_{A_{k}}$ of modes $A_{k}$, for any $k$, condensate in the
vacuum state which thus can be taken to be the memory state at
$t_{0}=0$ and which we denote by $|0>_{\cal N}$ (Vitiello 1995;
Alfinito and Vitiello, 2000). ${\cal N}_{A_{k}}(t)$ turns out to
be given, at each $t$, by:
\be
{\cal N}_{A_{k}}(t) \equiv {}_{\cal N}< 0(t) | A_{k}^{\dagger}
A_{ k} | 0(t) >_{\cal N} =  \sinh^{2}\bigl( \Gamma_{ k} t -
\theta_{k} \bigr) \quad , \lab{p274}
\ee
and similarly for the modes  ${\tilde A}_{k}$. The state
$|0(t)>_{\cal N} \equiv |0(\theta, t)>$ is the time-evoluted of
the state $|0>_{\cal N}$. $\Gamma$ is the damping constant
(related to the memory life-time) and $\theta_{k}$ fixes the code
value at $t_{0}=0$. $|0>_{\cal N}$ and $|0(t)>_{\cal N}$ are
normalized to $1$ and in the infinite volume limit we have
\be {}_{\cal N}<0(t) | 0>_{\cal N'}
   {\mapbelow{V \rightarrow \infty}} 0 \quad
\forall
  \, t \neq t_{0}, ~~\forall~ {\cal N}~, {\cal N'}~,
  \lab{p713}\ee
\be {}_{\cal N}<0(t) | 0(t')>_{\cal N'}
   {\mapbelow{V \rightarrow \infty}} 0,
  ~ ~\forall \, t\, , t'\, with \quad t \neq
  t'~, ~\forall ~{\cal N}~, {\cal N'}~ ,
\lab{p714} \ee
with $| 0(t)>_{\cal N'} \equiv | 0(\theta', t) > $. Eqs.
(\ref{p713}) and (\ref{p714}) also hold for ${\cal N} \neq {\cal
N'}$ but $t = t_{0}$ and $t = t'$, respectively.  Eqs.
(\ref{p713}) and (\ref{p714}) show that in the infinite volume
limit the vacua of the same code ${\cal N}$ at different times $t$
and  $t'$, for any $t$ and $t'$, and, similarly, at equal times
but different ${\cal N}$'s, are orthogonal states and thus the
corresponding Hilbert spaces are unitarily inequivalent spaces.

The number $\left( {\cal N}_{A_{k}} - {\cal N}_{{\tilde A}_{k}}
  \right)\,$ is a constant of motion for any $ k$ and  $\theta$.
The physical meaning of the ${\tilde A}$ system is the one of
providing the representation of the sink where the energy
dissipated by the $A$ system flows. Thermal properties of the
vacuum $| 0(t) >_{\cal N}$ can be then analyzed and the $\tilde A$
modes appear to represent the thermal bath (the environment)
modes.

In order to ensure the balance of energy flow between the system
and the environment,  the difference between the number of tilde
and non-tilde modes must be zero : ${\cal N}_{A_{k}} - {\cal
N}_{{\tilde A}_{k}} = 0$, for any $k$. Note that the requirement
${\cal N}_{A_{k}} - {\cal N}_{{\tilde A}_{k}} = 0, $ for any $ k$,
does not uniquely fix the code ${\cal N} \equiv \{{\cal
N}_{A_{k}}, ~for~any~  k \}$. Also ${|0>}_{\cal N'}$ with ${\cal
N'} \equiv \{ {\cal N'}_{A_{k}};  {\cal N'}_{A_{k}} - {\cal
N'}_{{\tilde A}_{k}} = 0,  ~for~any~ k \}$ ensures the energy flow
balance and therefore also ${|0>}_{\cal N'}$ is an available
memory state: it will correspond however to a different code
number  $(i.e.  {\cal N'})$  and therefore to a different
information than the one of code ${\cal N}$. In the infinite
volume limit  $\{{|0>}_{\cal N} \}$ and $\{{|0
>}_{\cal N'}\}$ are representations of the canonical comutation
relations each other unitarily inequivalent for different codes
${\cal N} \neq {\cal N'}$. We have thus at $t_{0} = 0$ the
splitting, or {\it foliation}, of the space of states into
infinitely many unitarily inequivalent representations. Thus,
infinitely many memory (vacuum) states, each one  of them
corresponding to a different  code $\cal N$, may exist: A huge
number of sequentially recorded inputs may  {\it coexist} without
destructive interference since infinitely many vacua  ${|0>}_{\cal
N}$, for all $\cal N$,  are {\it independently} accessible   in
the sequential recording process.

In conclusion, the "brain (ground) state" is represented as the
collection (or the superposition) of the full set of states
${|0>}_{\cal N}$, for all $\cal N$. The brain is thus described as
a complex system with a huge number of macroscopic states (the
memory states).

\vspace{1cm}

 {\it Part 3}

\vspace{0.3cm}

By considering the propagation speed, say $c$, of the NG modes in
the system, we can show that the time derivative with respect to
$t$ of the frequency $\Omega_{k}$ common to the $A$ and ${\tilde
A}$ modes, i.e. the power, is a decreasing function of $k$.
Similarly, we find that the inverse of $\Omega_{k}$ (the
''duration'') and  the domain size $d_{\Omega}(t) = c
(\Omega_{k})^{-1}$ are also decreasing functions of $k$.

It is possible to show that the degree of the coupling of the
system $A$ with the system ${\tilde A}$ can be parameterized by an
index, say $n$, in such a way that in the limit of $n \rightarrow
\infty$ the possibilities of the system $A$ to couple to ${\tilde
A}$ (the environment) are "saturated": the system $A$ then gets
{\it fully} coupled to $\tilde A$. $n$ can be thus taken to
represent the number of {\it links} between $A$ and ${\tilde A}$.
When $n$ is not very large (infinity), the system $A$ (the brain)
has not fulfilled its capability to establish links with the
external world (Alfinito and Vitiello, 2000). It can be shown that
more the system is "open" to the external world (more are the
links), better its neuronal correlation can be realized. However,
in the setting up of these correlations also enter quantities
which are intrinsic to the system, they are {\it internal}
parameters and may represent (parameterize) subjective attitudes.
Our model, however, is not able to provide a dynamics for the
variations of $n$, thus we cannot say if and how and under which
specific boundary conditions $n$ increases or decreases in time.
In any case, a higher or lower {\it degree of openness} (measured
by $n$) to the external world may produce a better or worse
ability in setting up neuronal correlates, respectively (different
under different circumstances, and so on, e.g. during the sleep or
the awake states, the childhood or the older ages).

In conclusion, functional or effective connectivity (as opposed to
the structural or anatomical one which we do not consider here) is
highly dynamic in the dissipative model. Once these functional
connections are formed, they are not necessarily fixed. On the
contrary, they may quickly change and new configurations of
connections may be formed extending over a domain including a
larger or a smaller number of neurons. The finiteness of the
correlated domain size implies a non-zero effective mass of the
DWQ. These therefore propagate through the domain with a greater
"inertia" than in the case of large (infinite) volume where they
are (quasi-)massless. The domain correlations are consequently
established with a certain time-delay. This concurs in the delay
observed in the recruitment of neurons in a correlated assembly
under the action of an external stimulus.

\vspace{1cm}

 {\it Part 4}

\vspace{0.3cm}

The free energy functional fot the system $A$ is
\be {\cal F}_{A} \equiv {}_{\cal N}<0(t)| \Bigl ( H_{A}
-{1\over{\beta}} S_{A} \Bigr ) |0(t)>_{\cal N}  \quad , \lab{(20)}
\ee
with the time-dependent inverse temperature $ {\beta (t) =
{1\over{k_{B} T(t)}}}$; $S_{A}$ is the entropy operator and
$H_{A}$ denotes the Hamiltonian at $t = t_{0}$ relative to the
$A$-modes only, ${H_{A} = \sum_{k} \hbar \Omega_{k}(t_{0})
A_{k}^{\dagger} A_{k}}$. Let $\Theta_{k} \equiv \Gamma_{k} t -
\theta_{k}$ and $E_{k} \equiv \hbar \Omega_{k}(t_{0})$. The
stationarity condition to be satisfied at each time $t$ by the
state $|0(t)>_{\cal N} $ is
\be {{\partial {\cal F}_{A}}\over{\partial \Theta_{k}}} = 0 \quad
, \quad \forall k \quad , \lab{(21)} \ee
and gives $ \beta (t) E_{k} = - \ln \tanh^{2} (\Theta_{k})$, i.e.
\be {\cal N}_{A_{k}}(\theta,t) = \sinh^{2} \bigl ( \Gamma_{k} t -
{\theta}_{k}\bigr ) = {1\over{{\rm e}^{\beta (t) E_{k}} - 1}}
\quad , \lab{(22)} \ee
which is the Bose distribution for $A_{k}$ at time $t$.

One can see that the entropy  ${\cal S}(t) = <0(t)| S |0(t)>_{\cal
N}$ is a decreasing function of time in the interval $(t_{0} = 0,
\tau \equiv \frac{\theta_{k}}{\Gamma_{k}} )$ meaning that the
state $|0(t)>_{\cal N}$, although evolving in time, is however
"protected" from "going back" to the "uncorrelated" vacuum state.
Of course, here it is crucial the energy exchange with the
environment and we are also assuming finite volume effects.
  One can also see that the entropy,
for both $A$ and $\tilde A$ system, grows monotonically from $0$
to infinity as the time goes  from   $t = \tau$ to $t = \infty$ .
However, for the complete system $A-{\tilde A}$, the difference
~$(S_{A} - S_{\tilde A})$~ is constant in time: $ [\, S_{A} -
S_{\tilde A} , {\cal H}^{\prime} ] = 0$.

Also, it can be shown that, as time evolves, the change in the
energy $ {E_{A} \equiv \sum_{k} E_{k} {\cal N}_{A_{k}}}$  and in
the entropy is given by
\be d E_{A} = \sum_{k} E_{k} \dot{\cal N}_{A_{k}} d t =
 {1\over{\beta}} d {\cal S}_{A}  \quad ,
  \lab{(23)}
\ee i.e. \be d E_{A} - {1\over{\beta}} d {\cal S}_{A} = 0 \quad ,
\lab{(24)} \ee
provided changes in inverse temperature are slow, i.e. $
{{{\partial \beta}\over{\partial t}} = - {1\over{k_{\tilde A}
T^{2}}} {{\partial T}\over{\partial t}} \approx ~0}$. In this
case, Eq. (\ref{(24)}) expresses the minimization of the free
energy: $ d {\cal F}_{A} = d E_{A} - {1\over{\beta}} d {\cal
S}_{A} = 0 $. One may define as usual heat as $
{dQ={1\over{\beta}} dS}$. Thus the change in time of condensate
(Eq. (\ref{(23)})) turns out into heat dissipation $dQ$.

\vspace{1cm}

 {\it Part 5}

 \vspace{0.3cm}

As already observed, the  state ${|0 (t)\rangle}_{\cal N}$  is a
normalized state at any $t$. Moreover, in the infinite volume
limit Eqs. (\ref{p713}) and (\ref{p714}) also hold true for ${\cal
N} = {\cal N'}$. Time evolution of the  state $|0\rangle_{\cal N}$
is thus represented as the (continuous) transition through the
representations $\{| 0(t) \rangle_{\cal N}, ~~\forall {\cal N}~,
~~\forall t \}$, namely by the ``trajectory" through the ``points"
$\{| 0(t) \rangle_{\cal N}, ~~\forall {\cal N}~,~~\forall t \}$ in
the space of the representations (each one minimizing the free
energy functional (\ref{(20)})).  The trajectory initial condition
at $t_{0} = 0$ is specified by the $\cal N$-set. It has been shown
(Vitiello, 2004; Pessa and Vitiello, 2003; 2004) that: {\it a)}
these trajectories are classical trajectories and {\it b)} they
are chaotic trajectories. This means that they satisfy the
requirements characterizing the chaotic behavior (Hilborn, 1994):

$i)$~ the trajectories are bounded and each trajectory does not
intersect itself (trajectories are not periodic).

$ii)$~~there are no intersections between trajectories specified
by different initial conditions.

$iii)$ trajectories of different initial conditions are
 diverging trajectories.

The meaning of $i)$ is that the ``points" $|0(t)\rangle_{\cal N}$
and $|0(t')\rangle_{\cal N}$ through which the trajectory goes,
for any $t$ and $t'$, with $t \neq t'$, after the initial time
$t_{0} = 0$, never coincide.

Eq. (\ref{p714}) also holds for ${\cal N} \neq {\cal N'}$ in the
infinite volume limit for any $t$ and any $t'$. Thus it shows that
trajectories specified by different initial conditions (${\cal N}
\neq {\cal N'}$) never cross each other, which is the meaning of
$ii)$.

The property $ii)$ thus implies that no {\it confusion}
(interference) arises among the codes of different neuronal
correlates, even as time evolves. In realistic situations of
finite volume, states with different codes may have non--zero
overlap (the inner products Eqs. (\ref{p713}) and (\ref{p714}) are
not zero). In such a case, at a ``crossing" point between two, or
more than two, trajectories, there can be ``ambiguities" in the
sense that one can switch from one of these trajectories to
another one which there crosses. This may be felt indeed as an
association of memories or as switching from one information to
another one and it reminds us of the ``mental switch" occurring,
for instance, during the perception of ambiguous figures and, in
general, while performing some perceptual and motor tasks as well
as while resorting to free associations in memory tasks (Eysenck,
1994).

In order to see that the requirement $iii)$ is also satisfied we
study how the ``distance"  between trajectories evolves as time
evolves. Consider two trajectories of different initial
conditions, ${\cal N} \neq {\cal N'}$ ($\theta \neq {\theta}'$).
At time $t$, each component ${\cal N}_{A_{k}}(t)$ of the code
${\cal N} \equiv \{ {\cal N}_{A_{k}} = {\cal N}_{{\tilde A}_{k}},
\forall k, at~~ t_{0}=0 \}$ is given by Eq. (\ref{p274}). We then
have:
$$
\Delta {\cal N}_{A_{k}}(t) \equiv {\cal
N'}_{A_{k}}(\theta',t) - {\cal N}_{A_{k}}(t) =
$$
\be\lab{co1} = \sinh^{2}\bigl ( \Gamma_{k} t - {\theta}_{k} +
{\delta \theta} \bigr ) - \sinh^{2}\bigl ( \Gamma_{k} t -
{\theta}_{k} \bigr ) \approx \sinh \bigl ( 2(\Gamma_{k} t -
{\theta}_{k}) \bigr ){\delta \theta_{k}} ~, \ee
where ${\delta \theta_{k}} \equiv {\theta}_{k} - {\theta}'_{k}$
(which, in full generality, may be assumed to be greater than
zero). The last equality holds for small ${\delta \theta_{k}}$,
i.e. for a very small difference in the initial conditions of the
two initial states. The time-derivative then gives
\be\lab{co3} \frac{\partial}{\partial t}\Delta {\cal N}_{A_{k}}(t)
= 2 {\Gamma_{k}} \cosh \bigl ( 2(\Gamma_{k} t - {\theta}_{k})
\bigr ){\delta \theta_{k}} ~. \ee
thus showing that the difference between originally even slightly
different ${\cal N}_{A_{k}}$'s grows as time evolves.  For large
enough $t$, the modulus of the difference $\Delta {\cal
N}_{A_{k}}(t)$ and its time derivative diverge as
$\exp{(2\Gamma_{k} t) }$, for all $k$'s. The quantity $ 2
\Gamma_{k}$, for each $k$, appears thus to play a role similar to
that of the Lyapunov exponent in chaos theory (Hilborn, 1994). In
conclusion, we see that trajectories differing by a small
variation $\delta \theta$ in the initial conditions, diverge
exponentially as time evolves. This may account for the high
perceptive resolution in the recognition of the perceptual inputs.

The difference between $k$--{\it components} of the codes $\cal N$
and $\cal N'$ may become zero at a given time $t_{k} =
\frac{\theta_{k}}{\Gamma_{k}}$ (cf. Eq. (\ref{co1})). However, the
difference between the codes $\cal N$ and $\cal N'$ does not
necessarily become zero. The codes  are different even if a finite
number of their components are equal since they are made up by a
large number of ${\cal N}_{A_{k}}(\theta,t)$ components (infinite
in the continuum limit). On the other hand, suppose that, for
${\delta \theta_{k}} \equiv {\theta}_{k} - {\theta'}_{k}$ very
small,  the time interval $\Delta t = \tau_{max} -\tau_{min}$,
with $\tau_{min}$ and $\tau_{max}$ the minimum and the maximum,
respectively, of $t_{k} = \frac{\theta_{k}}{\Gamma_{k}}$, for {\it
all} $k$'s, be very small. Then the codes are recognized to be
{\it almost} equal in such a $\Delta t$.  Eq. (\ref{co1}) then
expresses the recognition (or recall) process and it shows how it
is possible that ``slightly different'' ${\cal
N}_{A_{k}}$--patterns (or codes) are identified (recognized to be
the {\it same code} even if corresponding to slightly different
inputs). Roughly, $\Delta t$ may be taken as a measure of the
recognition time.

In conclusion, trajectories in the representation space are
classical chaotic trajectories in the large (infinite) volume
limit.

\newpage

\large{\bf References}

\vspace{0.5cm}

\normalsize
\begin{itemize}

\item[]
Alfinito, E. and G. Vitiello (2000). Formation and life--time of
memory domains in the dissipative quantum model of brain, Int. J.
Mod. Phys. B14, 853-868.

%\item[]
Alfinito, E. and G. Vitiello (2002). Domain formation in
noninstantaneous symmetry-breaking phase transitions, Phys. Rev. B
65, 054105.

Atmanspacher H, Scheingraber H. (1990) Pragmatic information and
dynamical instabilities in a multimode continuous-wave dye laser.
Can. J. Phys. 68: 728-737.

Bach-y-Rita P (1995) Nonsynaptic Diffusion Neurotransmission and
Late Brain Reorganization. New York: Demos-Vermande.

Bach-y-Rita P (2004) Tactile sensory substitution studies. Ann.
N.Y. Acad. Sci 1013: 83-91.

Bach-y-Rita P (2005) Emerging concepts of brain function. J Integr
Neurosci 4: 183-205.

Bak P (1996) How Nature Works: The Science of Self-organized
Criticality. New York: Copernicus.

Barrie JM. Freeman WJ and Lenhart M (1996) Modulation by
discriminative training of spatial patterns of gamma EEG amplitude
and phase in neocortex of rabbits. J Neurophysiol 76: 520-539.

Bartlett FC (1932) Remembering. Cambridge University Press.

Basar E (1998) Brain Function and Oscillations. Berlin:
Springer-Verlag.

Becker CJ and Freeman WJ (1968) Prepyriform electrical activity
after loss of peripheral or central input or both. Physiol Behav
3: 597-599.

Braitenberg V, Sch\"uz A (1991) Anatomy of the Cortex: Statistics
and Geometry. Berlin: Springer-Verlag.

Bratteli O. and D. W. Robinson (1979). Operator Algebra and
Quantum Statistical Mechanics. Berlin: Springer.

Bressler SL, Kelso JAS. (2001). Cortical coordination dynamics and
cognition. Trends Cog. Sci. 5:26-36.

Bullock TH (1959) The neuron doctrine and electrophysiology.
Science 129: 997-1002.

Burns BD (1958) The Mammalian Cerebral Cortex. London: Arnold.

Celeghini, E., M. Rasetti and G. Vitiello (1992). Quantum
dissipation, Ann. Phys. 215, 156-170.

Cohen LG, Celnik P, Pascal-Leone A, Corwell B, Faiz L, Dambrosia
J, Honda M, Sadato N, Gerloff C, Catala MD, Hallett M (1997)
Functional relevance of cross-modal plasticity in blind humans.
Nature 389: 180 - 183; doi:10.1038/38278.

Davia, C. J. (2005) Life, catalysis and excitable media: A dynamic
systems approach to metabolism and cognition. In J. Tuszynski
(Ed.)  The Emerging Physics of Consciousness.  Heidelberg,
Germany: Springer-Verlag.

De Concini, C. and G. Vitiello (1976). Spontaneous breakdown of
symmetry and group contraction. Nucl. Phys. B116, 141-156.

Del Giudice, E., S. Doglia, M. Milani and G. Vitiello (1985). A
quantum field theoretical approach to the collective behavior of
biological systems. Nucl. Phys. B251 (FS 13), 375-400.

Del Giudice, E., S. Doglia, M. Milani and G.Vitiello (1986).
Electromagnetic field and spontaneous symmetry breakdown in
biological matter. Nucl. Phys. B275 (FS 17), 185-199.

Del Giudice, E., G. Preparata and G. Vitiello (1988). Water as a
free electron laser. Phys. Rev. Lett. 61, 1085- 1088.

Dunn J. R., Fuller M., Zoeger J., Dobson J., Heller F., Hammann
J., Caine E., and Moskowitz B. M. Magnetic material in the human
hippocampus. Brain Re.s Bull. 36 . (1995) pp. 149-153.

Elul R (1972) The genesis of the EEG. Int Rev Neurobiol 15:
227-272.

Eysenck, M.W. (1994). Principles of Cognitive Psychology, Erlbaum,
Hillsdale, NJ, 1994.

Fingelkurts AA, Fingelkurts AA. (2001) Operational architectonics
of the human brain biopotential field: toward solving the
mind-brain problem. Brain and Mind 2: 261-296.

Fingelkurts AA, Fingelkurts AA (2004) Making complexity simpler:
multivariability and metastability in the brain. Int J Neurosci
114: 843-862.

Franken P, Malafosse A, Tafti M (1998) Genetic variation in EEG
activity during sleep in inbred mice. Am. J. Physiol. 275:
R1127-1137.

Freeman W.J.. (1967) Analysis of function of cerebral cortex by
use of control systems theory.  The Logistics Review 3: 5-40.

Freeman W.J.. (1975/2004) Mass Action in the Nervous System. New
York: Academic Press;
http://sulcus.berkeley.edu/MANSWWW/MANSWWW.html

Freeman W.J. (1986) Petit mal seizure spikes in olfactory bulb and
cortex caused by runaway inhibition after exhaustion of
excitation. Brain Research Reviews 11:259-284.

Freeman W.J. (1990) On the problem of anomalous dispersion in
chaoto-chaotic phase transitions of neural masses, and its
significance for the management of perceptual information in
brains.  Chapter in: Haken H, Stadler M (eds.) Synergetics of
Cognition. Berlin, Springer-Verlag, Vol 45: 126-143.

Freeman WJ (1998) Bidirectional processing in the olfactory-limbic
axis during olfactory behavior. Behavioral Neuroscience 112:
541-553.

Freeman W.J.. (2000) Neurodynamics. An Exploration of Mesoscopic
Brain Dynamics. London UK: Springer-Verlag.

Freeman W.J. (2001) How Brains Make Up Their Minds. New York:
Columbia University Press.

Freeman W.J. (2004a) Origin, structure, and role of background EEG
activity. Part 1. Phase. Clin. Neurophysiol. 115: 2077-2088.

Freeman W.J. (2004b) Origin, structure, and role of background EEG
activity. Part 2. Amplitude. Clin. Neurophysiol. 115: 2089-2107.

Freeman W.J. (2005a) Origin, structure, and role of background EEG
activity. Part 3. Neural frame classification. Clin. Neurophysiol.
116: 1117-1129.

Freeman WJ (2005b) Origin, structure, and role of background EEG
activity. Part 4. Neural frame simulation. Clin Neurophysiol, in
press.

Freeman WJ (2005c) A field-theoretic approach to understanding
scale-free neocortical dynamics. Biol Cybern 92/6: 350-359.

Freeman WJ (2006) A cinematographic hypothesis of cortical
dynamics in perception. Intern J Psychophysiol, in press.

Freeman W.J., Baird, B. (1989) Effects of applied electric current
fields on cortical neural activity.  Chapter in: Schwartz E (ed.)
Computational Neuroscience.  New York, Plenum Press.  pp. 274-287.

Freeman WJ, Barrie JM (2000) Analysis of spatial patterns of phase
in neocortical gamma EEGs in rabbit. Journal of Neurophysiology
84: 1266-1278.

Freeman W.J., Burke BC (2003) A neurobiological theory of meaning
in perception.  Part 4. Multicortical patterns of amplitude
modulation in gamma EEG. Int. J. Bifurc. Chaos 13: 2857-2866.

Freeman W.J., Burke BC, Holmes MD (2003) Aperiodic phase
re-setting in scalp EEG of beta-gamma oscillations by state
transitions at alpha-theta rates. Human Brain Mapping
19(4):248-272.

Freeman W.J., Burke BC, Holmes MD, Vanhatalo S (2003) Spatial
spectra of scalp EEG and EMG from awake humans. Clin.
Neurophysiol. 114: 1055-1060.

Freeman W.J., Ga\'al G, Jornten R (2003) A neurobiological theory
of meaning in perception.  Part 3. Multiple cortical areas
synchronize without loss of local autonomy. Int. J. Bifurc. Chaos
13: 2845-2856.

Freeman WJ, Grajski KA (1987) Relation of olfactory EEG to
behavior: Factor analysis: Behav. Neurosci.   101: 766-777.

Freeman W.J., Holmes MD, West GA, Vanhatalo S (2005) Dynamics of
human neocortex that optimize its stability and flexibility. J.
Intelligent. Syst., in press.

Freeman, W.J. and Rogers, L.J. (2003) A neurobiological theory of
meaning in perception.  Part 5. Multicortical patterns of phase
modulation in gamma EEG.  Int. J. Bifurc. Chaos 13: 2867-2887.

Friston KJ (2000) The labile brain. I. Neuronal transients and
nonlinear coupling. Phil Trans R Soc Lond B 355:215-236.

Fuxe, K., Agnati, L. F. (1991) Volume transmission in the Brain.
New York: Raven Press.

Gray CM and Skinner JE (1988) Centrifugal regulation of neuronal
activity in the olfactory bulb of the waking rabbit as revealed by
reversible cryogenic blockade. Exp Brain Res 69:378-386.

Haken H (1983) Synergetics: An Introduction. Berlin:
Springer-Verlag. Hilborn R. (1994) Chaos and nonlinear Dynamics.
Oxford: Oxford University Press,

Haken H (1996) Principles of Brain Functioning: A Synergetic
Approach to Brain Activity, Behavior, and Cognition. New York:
Springer.

Haken H (2004) Synergetics: Introduction and Advanced Topics. New
York: Springer.

Hawkins J (2004) On Intelligence. With Blakeslee S. New York:
Times Books.

Hilborn R (1994) Chaos and nonlinear Dynamics. Oxford: Oxford
University Press,

Hoppensteadt FC and Izhkevich EM (1998) Thalamo-cortical
interactions modeled by weakly connected oscillators: could the
brain use FM radio principles?  BioSystems 48: 85-94.

Houk JC (2001) Neurophysiology of frontal-subcortical loops. Ch. 4
in: Lichter DG, Cummings JL (eds.) Frontal-Subcortical Circuits in
Psychiatry and Neurology. New York: Guilford Publ., pp. 92-113.

Hwa R.C. and Ferree T. (2002) Scaling properties of fluctuations
in the human electroencephalogram. Physical Rev. , E 66: 021901.

Ingber L (1995) Statistical mechanics of multiple scales of
neocortical interactions. In: Nunez PL (ed.) Neocortical Dynamics
and Human EEG Rhythms. New York: Oxford UP, pp. 628-681.

Jensen H.J. (1998) Self-Organized Criticality: Emergent Complex
Behavior in Physical and Biological Systems.  New York: Cambridge
UP.

Jibu, M. and K. Yasue (1992). A physical picture of Umezawa's
quantum brain dynamics. In R. Trappl (Ed.), Cybernetics and System
Research, p.797-804. Singapore: World Scientific.

Jibu, M., and K. Yasue (1995). Quantum brain dynamics and
consciousness. Amsterdam: John Benjamins. Kay LM,

Kay LM, Freeman WJ (1998) Bidirectional processing in the
olfactory-limbic axis during olfactory behavior. Behav Neurosci
112: 541-553.

Kelso, J.A.S. (1995). Dynamic Patterns: The Self Organization of
Brain and Behavior. Cambridge: MIT Press.

K\"ohler W (1940) Dynamics in Psychology. New York NY: Grove
Press.

K\"onig P and Schillen TB (1991) Stimulus-dependent assembly
formation of oscillatory responses: I. Synchronization. Neural
Computation 3: 155-166.

Kozma R, Freeman W.J. (2001): Chaotic resonance: Methods and
applications for robust classification of noisy and variable
patterns. Int J Bifurc Chaos 10: 2307-2322.

Kozma R, Freeman W.J., Erd\'i P (2003): The KIV model - nonlinear
spatio-temporal dynamics of the primordial vertebrate forebrain.
Neurocomputing 52:819-826.

Kozma R, Puljic M, Balister P, Bollobás B, Freeman W.J. (2005)
Emergence of collective dynamics in the percolation model of
neural populations: Mixed model with local and non-local
interactions. Biol. Cybern., in press.

Lashley KS (1929) Brain Mechanisms and Intelligence. Chicago: U.P.
Chicago.

Lashley KS (1942) The problem of cerebral organization in vision.
In J Cattell (Ed.), Biological Symposia VII: 301-322.

Li G, Lou Z, Wang L, Li X and Freeman WJ. (2005) Application of
chaotic neural model based on olfactory system on pattern
recognitions. pp. 378-381 in: Wang L, Chen K, Ong YS (eds.) ICNC
2005, LNCS 3610, Berlin: Springer.

Li X, Li G, Wang L and Freeman WJ (2005) A study on a bionic
pattern classifier based on olfactory neural system. Intern J.
Bifurc. Chaos, in press.

Linkenkaer-Hansen K, Nikouline VM, Palva, JM, Iimoniemi RJ. (2001)
Long-range temporal correlations and scaling behavior in human
brain oscillations.  J Neurosci, 15: 1370-1377.

Lyamin OI, Mukhametov IM, Siegel JM, Nazarenko EA, Polyakova IG,
Shpak OV. (2002) Unihemispheric slow wave sleep and the state of
the eyes in a white whale. Behav. Brain Res. 129: 125-129.

Maturana H R., Varela F J (1980) Autopoiesis and Cognition: The
Realization of the Living (Boston Studies in the Philosophy of
Science).  Berlin: Springer-Verlag. Merleau-Ponty M (1945/1962)
Phenomenology of Perception. (C Smith, Trans.). New York:
Humanities Press.

Merleau-Ponty M (1945/1962) Phenomenology of Perception. (C Smith,
trans.). New York: Humanities Press.

Miller LM and Schreiner CE (2000) Stimulus-based state control in
the thalamocortical system. J Neurosci 20:7011-7016.

Nunez PL (1981) Electric Fields of the Brain: The Neurophysics of
EEG. New York: Oxford UP.

Ohl FW, Scheich H, Freeman W.J.. (2001) Change in pattern of
ongoing cortical activity with auditory category learning.  Nature
412:733-736.

Ohl FW, Scheich H and Freeman WJ (2005) Neurodynamics in auditory
cortex during category learning. Ch. 8 in: K\"onig R, Heil P,
Budinger E, Scheich H (eds.) The Auditory Cortex - A Synthesis of
Human and Animal Research. Mahwah NJ: Lawrence Erlbaum, pp.
429-444.

Orlovskii GN, Deliagina TG and Grillner S (1999) Neuronal Control
of Locomotion. New York: Oxford UP.

Pessa, E. and G. Vitiello (2003). Quantum noise, entanglement and
chaos in the quantum field theory of mind/brain states. Mind and
Matter 1, 59--79.

Pessa, E. and G. Vitiello (2004). Quantum noise induced
entanglement and chaos in the dissipative quantum model of brain.
Int. J. Mod. Phys. B18, 841-858.

Piaget J (1930)  The child's conception of physical causality.
New York: Harcourt, Brace. Pikovsky A, Rosenblum M, Kurths J.
(2001) Synchronization - A Universal Concept in Non-linear
Sciences.  Cambridge UK: Cambridge U.P.

Pikovsky A, Rosenblum M and Kurths J (2001) Synchronization - A
Universal Concept in Non-linear Sciences.  Cambridge UK: Cambridge
UP.

Pribram KH (1971) Languages of the Brain. Engelwood Cliffs NJ:
Prentice-Hall.

Prigogine I (1980) From Being to Becoming: Time and Complexity in
the Physical Sciences. San Francisco: W H Freeman.

Ricciardi, L. M. and H. Umezawa (1967). Brain and physics of
many-body problems, Kybernetik 4, 44--48. Reprint in Brain and
Being, G. G. Globus, K. H. Pribram and G. Vitiello Eds. p.
255-266. Amsterdam: John Benjamins.

Robinson PA, Rennie CJ, Rowe DL and O'Connor SC (2004) Estimation
of multiscale neurophysiologic parameters by
electroencephalographic means. Hum Brain Map 23: 53-72.

Roelfsema PR, Engel AK, K\"onig P and Singer W (1997) Visuomotor
integration is associated with zero time-lag synchronization among
cortical areas. Nature 385: 157-61.

Shimoide K, Freeman W. J. (1995) Dynamic neural network derived
from the olfactory system with examples of applications.  IEICE
Transaction Fundamentals E-78A: 869-884.

Singer W, Gray CM (1995) Visual feature integration and the
temporal correlation hypothesis.  Ann Rev Neurosci 18: 555-586.

Steriade M (1997) Synchronized activities in coupled oscillators
in the cerebral cortex and thalamus at different levels of
vigilance. Cerebral Cortex 7:583-604.

Skarda CA, Freeman W.J. (1987) How brains make chaos in order to
make sense of the world. Brain and Behavioral Science 10: 161-195.

Stuart, C. I. J., Y. Takahashi and H. Umezawa (1978). On the
stability and non-local properties of memory, J. Theor. Biol. 71,
605--618.

Stuart, C. I. J., Y. Takahashi and H. Umezawa (1979). Mixed system
brain dynamics: neural memory as a macroscopic ordered state,
Found. Phys. 9, 301--327.

Traub RD, Whittington MA, Stanford IM and Jefferys JGR (1996) A
mechanism for generation of long-range synchronous fast
oscillations in the cortex. Nature 383: 421-424.

Tsuda I. (2001) Towards an interpretation of dynamic neural
activity in terms of chaotic dynamical systems. Behav Brain Sci
24:793-810.

Umezawa, H. (1993). Advanced field theory: micro, macro and
thermal concepts. New York: American Institute of Physics.

Umezawa, H., and G. Vitiello (1985). Quantum Mechanics. Naples:
Bibliopolis.

Vitiello, G. (1995). Dissipation and memory capacity in the
quantum brain model, Int. J. Mod. Phys. B9, 973--989.

Vitiello, G. (2001). My Double Unveiled. Amsterdam: John
Benjamins.

Vitiello, G. (2004). Classical chaotic trajectories in quantum
field theory. Int. J. Mod. Phys. B18, 785-792.

Von Melchner L., Pallas S.L., Sur M. (2000) Visual behaviour
mediated by retinal projections directed to the auditory pathway.
Nature 404: 871 - 876; doi:10.1038/35009102.

Wakeling J.R. (2004) Adaptivity and 'Per learning'. Physica A 340:
766-773.

Wang X.F., Chen G.R. (2003) Complex networks: small-world,
scale-free and beyond. IEEE Circuits Syst., 31: 6-20.

Watts D.J. and Strogatz S.H. (1998) Collective dynamics of
"small-world" networks. Nature 393: 440-442.

Whittington M.A., Faulkner H.J., Doheny H.C. and Traub R.D. (2000)
Neuronal fast oscillations as a target site for psychoactive
drugs. Pharmacol Therap 86: 171-190.

Wilson HR, Cowan J.D. (1973) Excitatory and inhibitory
interactions in localized populations of model neurons. Biophysics
Journal 12: 1-24.

Wright J.J., Alexander D.M. and Bourke PD (2005) Contribution of
lateral interactions in V1 to organization of response properties.
Vision Research, in press.

Wright J.J., Bourke P.D., Chapman C.L. (2000) Synchronous
oscillation in the cerebral cortex and object coherence:
simulation of basic electrophysiological findings. Biol. Cybern.
83: 341-353.

\end{itemize}

\end{document}